\documentclass[twocolumn, trackchanges]{aastex631}
\usepackage{amsmath}
\usepackage{graphicx}
\usepackage{epsf}
\usepackage{color}
\usepackage{enumitem}
\usepackage{csvsimple,booktabs}
\usepackage[flushleft]{threeparttable}

\begin{document}

\title{\textbf{A Search for Lensed Ly$\boldsymbol\alpha$ Emitters within the Early HETDEX Data Set}}
\author[0000-0003-4323-0597]{Isaac H. Laseter}
\affiliation{Department of Astronomy, University of Wisconsin-Madison, Madison, WI 53706}
\affiliation{Department of Astronomy, The University of Texas at Austin, 2515 Speedway Boulevard, Austin, TX 78712, USA}
\author[0000-0001-8519-1130]{Steven L. Finkelstein}
\affiliation{Department of Astronomy, The University of Texas at Austin, 2515 Speedway Boulevard, Austin, TX 78712, USA}
\author{Micaela J. Bagley}
\affiliation{Department of Astronomy, The University of Texas at Austin, 2515 Speedway Boulevard, Austin, TX 78712, USA}
\author{Dustin M. Davis}
\affiliation{Department of Astronomy, The University of Texas at Austin, 2515 Speedway Boulevard, Austin, TX 78712, USA}

\author[0000-0002-8433-8185]{Karl Gebhardt}
\affiliation{Department of Astronomy, The University of Texas at Austin, 2515 Speedway Boulevard, Austin, TX 78712, USA}

\author{Caryl Gronwall}
\affiliation{Department of Astronomy $\&$ Astrophysics, The Pennsylvania State University, University Park, PA 16802, USA}
\affiliation{Institute for Gravitation and the Cosmos, The Pennsylvania State University, University Park, PA 16802, USA}

\author[0000-0002-1328-0211]{Robin Ciardullo}
\affiliation{Department of Astronomy $\&$ Astrophysics, The Pennsylvania State University, University Park, PA 16802, USA}
\affiliation{Institute for Gravitation and the Cosmos, The Pennsylvania State University, University Park, PA 16802, USA}

\author{Gregory R. Zeimann}
\affiliation{Hobby Eberly Telescope, University of Texas, Austin, Austin, TX, 78712, USA}

\author[0000-0002-2307-0146]{Erin Mentuch Cooper}
\affiliation{Department of Astronomy, The University of Texas at Austin, 2515 Speedway Boulevard, Austin, TX 78712, USA}
\affiliation{McDonald Observatory, The University of Texas at Austin, Austin, TX 78712}

\author[0000-0003-2575-0652]{Daniel Farrow}
\affiliation{Max-Planck-Institut f{\"u}r extraterrestrische Physik, Giessenbachstrasse 1, 85748 Garching, Germany}
\affiliation{Universit{\"a}ts-Sternwarte, Fakult{\"a}t f{\"u}r Physik, Ludwig-Maximilians-Universit{\"a}t M{\"u}nchen, Scheinerstr. 1, 81679  M{\"u}nchen, Germany}

\begin{abstract}

The Hobby-Eberly Telescope Dark Energy Experiment (HETDEX) is a large volume spectroscopic survey without pre-selection of sources, searching $\sim 540$ deg$^2$ for Ly$\alpha$ emitting galaxies (LAEs) at $1.9 < z < 3.5$. Taking advantage of such a wide-volume survey, we perform a pilot study using early HETDEX data to search for lensed Ly$\alpha$ emitters (LAEs). After performing a proof-of-concept using a previously known lensed LAE covered by HETDEX, we perform a search for previously unknown lensed LAEs in the HETDEX spectroscopic sample. We present a catalog of 26 potential LAEs lensed by foreground, red, non-star-forming galaxies at $z \sim 0.4 - 0.7$. We estimate the magnification for each candidate system, finding $12$ candidates to be within the strong lensing regime (magnification $\mu > 2$). Follow-up observations of these potential lensed LAEs have the potential to confirm their lensed nature and explore these distant galaxies in more detail. 

\vspace{1cm}

\end{abstract}

\section{Introduction} \label{Intro}

Ly$\alpha$ Emitters (LAEs) have been pivotal to the current understanding of galaxy evolution and the epoch of reionziation. While LAEs are typically thought of as young, low-mass, star forming galaxies \citep{Partridge_1967}, it has been shown that LAEs exhibit a range of properties such as dust content, ages, masses, luminosities, and line profiles (e.g., red line offsets and multi-component structures with blue wings), leading to a complex and incomplete picture of these galaxies \citep{Gawiser_2006, Finkelstein_2006, Finkelstein_2008, Lai_2008, Finkelstein_2009, Pen_2009, Cowie_2011, Nilsson_2011, Guaita_2011, Atek_2014, Song_2014, Gre_2017, De_Barros_2017, Reddy_2021}. An important step in constraining the properties of LAEs is to constructing large samples of LAEs spanning a range of luminosities \citep[e.g.,][]{Cowie_1998,Finkelstein_2011, Shu_2016, Hill_2016, Herenz_2018, Cao_2020}.

There have been two primary ways LAE samples have been constructed: 1) Narrowband imaging surveys and 2) Spectroscopic surveys. Narrowband imaging has the advantage of using optical imagers which allow for large areas of the sky to be probed ($\rm \sim 1 deg^2$; e.g., the Hawaii eROSITA Ecliptic Pole Survey (HEROES, \citealt{Songaila_2018}), the Lyman Alpha Galaxies in the Epoch of Reionization (LAGER, \citealt{Zheng_2017}), the Large Area Lyman Alpha survey (LALA, \citealt{rhoads2001}), and the One-hundred-square-degree DECam Imaging in Narrowbands survey (ODIN, \citealt{Yun_2022}). However, narrowband surveys are limitated to narrow redshift ranges (e.g., a typical bandwidth is $\rm \sim \Delta z = 0.1$), and require spectroscopic followup to confirm candidates as LAEs. Spectroscopic surveys have been performed through low resolution slitless spectroscopy \citep{Kurk_2004, Galex_2005}, blank-sky slit spectroscopy \citep{Crampton_1999, Tran_2004, Rauch_2008, Sawicki_2008, Cassata_2011}, and integral-field spectroscopy \citep{Bacon_2010, Adams_2011}. Although spectroscopic surveys have the benefit of searching over a larger redshift range due to wider wavelength coverage, the total volume probed is limited due to the limited area of most spectrographs. Overall, a wide area, spectroscopic survey would be the most efficient way of building large samples of LAEs.

While narrowband imaging has historically been more widely used due to the availability of wide-field imagers, most narrowband searches have been limited to LAEs at L$^{\ast}_{Ly\alpha}$ or brighter, thus many of these imaging surveys have been unable to fully probe the faint end of the Ly$\alpha$ luminosity function. While recent deep spectroscopic surveys have begun to probe the faint-end \citep[e.g.,][]{De_La_Vieuville_2019}, the areas observed by such studies are small, and the faintest galaxies are observed at fairly low signal-to-noise, making it difficult to study their physical properties.

One avenue to study fainter LAEs at higher signal-to-noise is to locate LAEs which have been gravitationally lensed by foreground galaxies \citep[galaxy-galaxy lensing; e.g.,][]{Warren_1996, Maizy_2009, Shu_2016, De_La_Vieuville_2019} or by galaxy clusters \citep[cluster lensing; e.g.,][]{Atek_2015, Fuller_2020ApJ}. Galaxy-galaxy lensing surveys, such as the Sloan Lens Advanced Camera for Surveys (SLACS) \citep{Bolton_2006}, SLACS for the Masses Survey (S4TM) \citep{Shu_2015}, the Baryon Oscillation Spectroscopic Survey (BOSS) Emission-Line Lens Survey Galaxy-Ly$\alpha$ Emitter Systems (BELLS GALLERY) Survey \citep{Shu_2016}, and Lensed LAEs in the EBOSS Survey (LESSER) \citep{Lesser_2020}, have the advantage of identifying lensing systems through spectroscopic observations of the foreground galaxy, but are limited in magnification relative to cluster lensing due to the foreground mass. Cluster lensing surveys, such as the Cluster Lensing And Supernova survey with Hubble (CLASH) \citep{Postman_2012}, the Hubble Frontier Fields (HFF) \citep{Lotz_2016, Koekemoer_2016}, and the Reionization Lensing Cluster Survey (RELICS) \citep{Coe_2019}, have the benefit of pushing the observational limits of high redshift studies (z $\sim 10$), but typically require deep imaging (e.g., 190 Hubble Space Telescope orbits for RELICS) with spectroscopic follow-up.

Whether it is through galaxy-galaxy lensing or cluster lensing, identifying lensing systems with background LAEs enables fainter LAEs to be studied in detail due to magnification effects \citep{Stark_2007, Maizy_2009, Ota_2012, Mason_2015, Shu_2015, Shu_2016, Lesser_2020}. The potential level of ``detail” is dependent on the magnification; a higher magnification ($\rm \mu$) factor both increases the apparent brightness and also can result in higher apparent angular resolution. Therefore, it is desirable to isolate lensed LAEs in the strongest lensing regimes. If one targets cluster lenses strong gravitationally lensed LAEs are more common relative to galaxy-galaxy lensing \citep[e.g.,][]{Bina_2016, De_La_Vieuville_2019, Fuller_2020ApJ, Claeyssens_2022}, but obtaining the deep imaging and spectroscopic follow-up presents its own obstacles. For galaxy-galaxy lensing systems strong gravitationally lensed LAEs are rarer, and thus large volume surveys are required. In this work we target galaxy-galaxy lensed LAEs, and thus we require a large volume survey.
 
The Hobby-Eberly Telescope Dark Energy Experiment \citep[HETDEX;][]{Gebhardt_2021} is a wide-area, integral field spectroscopic survey which will ultimately cover 
$\sim$540 deg$^2$.
HETDEX covers 3500-5500 \AA, and is thus sensitive to Ly$\alpha$ emission at 1.9 $< z <$ 3.5, providing an excellent opportunity for a wide-area search for rare lenses.  Here we use an early catalog from this in-progress survey to perform a pilot search for LAEs lensed by foreground massive galaxies.  For this pilot search, we limit ourselves to red, presumably non-star-forming galaxies, such that any detected emission line should be from a background object, limiting contamination in our survey.

Due to the blue wavelengths observed by the HETDEX Spectrograph (see Section \ref{HETDEX}), detected background emission lines are probable to be Ly$\alpha$. Using the Ly$\alpha$ redshift, the redshift of the lensed galaxy, the angular distance between sources, and lens modeling, the magnification of the system can be calculated. In doing so, favorable systems can be identified for follow up observations to confirm the presence of LAEs, provide unparalleled details of fainter LAEs, and to aid in ongoing research where fainter LAEs are paramount.

This paper is organized in the following way. In Section \ref{Sec. 2} we discuss a ``proof of concept'' of searching for lensed LAEs within HETDEX and how we identified potential foreground lens galaxies to hunt for gravitationally lensed LAEs. In Section \ref{Building a Sample of Candidate Lensed LAEs} we present how we acquired spectra for each potential lensing system and define our procedures for identifying potential background LAEs. In Section \ref{Magnification Determination} we present the procedure of determining the magnification of identified LAE lensing systems, and finally in Section \ref{Summary} we summarize our results and discuss future work that can spur from this project. We use a H$_0$ = $70$ km/s/Mpc, $\rm \Omega_{m} = 0.3$, $\rm \Omega_{\lambda}$ $= 0.7$ cosmology throughout the paper.

\section{Proof of Concept} \label{Sec. 2}

\subsection{HETDEX} \label{HETDEX}

HETDEX is a spectroscopic survey aimed at measuring the Hubble parameter, $H(z)$, and the angular diameter distance, $D_A(z)$, between the redshifts $1.88 < z < 3.52$, the epoch when the majority of the stellar mass in the universe formed \citep{Madau_2014}, by observing $\sim$ 1 million LAEs. Hunting for LAEs within this redshift range is designed to measure the clustering of galaxies to constrain the evolution of the energy density of dark energy. However, the sheer number of LAEs observed through HETDEX provides an extensive dataset to explore other frontiers of astronomy such as galaxy evolution.

HETDEX uses the large Visible Integral-field Replicable Unit Spectrograph \citep[VIRUS;][]{Hill_2016, Indahl_2019, Hill_2021}, which covers $\rm 3,500-5,500 \AA$ with a resolution of $\rm R \sim 800$ using up to $78$ integral field units (IFUs) ($\rm 51\arcsec$ x $\rm 51\arcsec$ per IFU), each based on a similar design and each with differences in performance and build.  Together, the total number of usable fibers ($\rm 1.5\arcsec$ diameter) with VIRUS  is $\sim 34,000$.  By performing a spectroscopic survey over $\sim 540$ deg$^2$, the HETDEX spectroscopic sample size is comparable to the Sloan Digital Sky Survey (SDSS) spectroscopic sample, only at a time $2-3$ billion years after the Big Bang. HETDEX is ongoing, with $60\%$ of the survey completed by the end of 2021 \citep{Gebhardt_2021}.

The HETDEX reduction algorithm employs an automated emission$-$line$-$finding code, extracting fiber spectra to identify emission lines with no positional prior. The vast majority of detections consist of just a single emission line, which is likely to be either Ly$\alpha$ or [OII]. To classify these lines, HETDEX then employs multiple analyses which uses all fiber spectra, along with other information such as bandpass continuum estimates, physical size estimates, overlapping neighbor PSFs, etc.\ from imaging and multiwavelength photometry when available \citep{Leung_2017, Farrow_2021, Davis_2021}. The imaging utilized by the HETDEX emission$-$line$-$finding code involves a suite of imaging surveys: Hyper Suprime Cam - Dark Energy Experiment (HSC-DEX; HETDEX Specific Survey) ($m_{lim} \sim 25.5$ for a $2\arcsec$ aperture), the Dark Energy Camera Legacy Survey (DECaLS; $m_{lim} g\sim24.0$ and $m_{lim} r\sim23.4-24.0$ for a $2\arcsec$ aperture), and the HETDEX-Imaging Mosaic taken with the Kitt Peak National Observatory (KPNO) Mayall 4-meter telescope in the g-band ($m_{lim} g\sim23.4$ for a $2\arcsec$ aperture; observations taken in 2011, 2012, and 2015). HETDEX itself has a $m_{lim} \sim$ 24.5-25.0 for continuum detection when treating the HETDEX spectra as a bandpass. In short, the HETDEX  emission line finding code utilizes the HETDEX computed mag and every imaging limited magnitude, which all feed into making a single estimate for the magnitude. This estimated magnitude is then used as a proxy for the continuum, thus enabling a classification/line probability to be determined. Sections \ref{Extracting Spectra} -- \ref{Identification Pipeline/Analysis} explore this procedure in detail.

HETDEX began survey observations in 2017 and has since continued to increase the number and performance of VIRUS units and improve data handling. For the current pilot project, we use HETDEX internal data release 2 (HDR2), which encompasses over $\sim$ 210 million spectra, or roughly 35\% of the eventual fully completed dataset.

\subsection{Searching for Known Lensed LAEs} \label{Searching for Known Lensed LAEs}

As this work represents the first exploration of the complex HETDEX dataset for lensed LAEs, we first performed a ``proof of concept” study using the positions of 187 potential gravitationally lensed LAEs reported from the BELLS for the GALaxy-Ly$\alpha$ EmitteR sYstems Survey \citep[BELLS GALLERY;][]{Shu_2016}. The work by \cite{Shu_2016} was chosen for comparison due to their foreground galaxy selection and technique in identifying a lensed LAE candidate. 
Should any of their lensed sources be covered by HETDEX, we could then use those sources as a test of our ability to discover lensed sources with these data.

\cite{Shu_2016} utilized data from data release 12 of the Baryon Oscillation Spectroscopic Survey (BOSS), which had a primary goal of delivering redshifts of $\sim 1.5$ million luminous red galaxies out to $z = 0.7$ \citep{Dawson_2013}, which represent ideal locations to search for lensed LAEs. \cite{Shu_2016} identified potential lensed LAEs by parsing foreground spectra for ``rogue" emission lines originating from a background LAE; a similar procedure to the current work discussed in Section \ref{Building a Sample of Candidate Lensed LAEs}. The results from \cite{Shu_2016} were building off a history of successful SDSS lensing surveys such as SLACS \citep{Bolton_2006} and S4TM \citep{Shu_2015}. 

To explore which, if any, BELLS GALLERY systems were observed by HETDEX, we searched the HDR2 database for all HETDEX detections within a radius of $360\arcsec$ of the positions of the known lensed LAEs. The reasoning for such a sizable search radius is to ensure the BELLS GALLERY lensed candidate positions approaching the edge of a HETDEX pointing would be included in this sample. For any matches, we then located and extracted the closest fiber to the positions of the potential lensed LAEs within the HETDEX/VIRUS pointings we returned. We then visually inspected the returned fiber spectra to ensure the respective potential lensed LAEs were encompassed within the fiber coverage.

\begin{figure*}
    \centering
    \includegraphics[width = \textwidth]{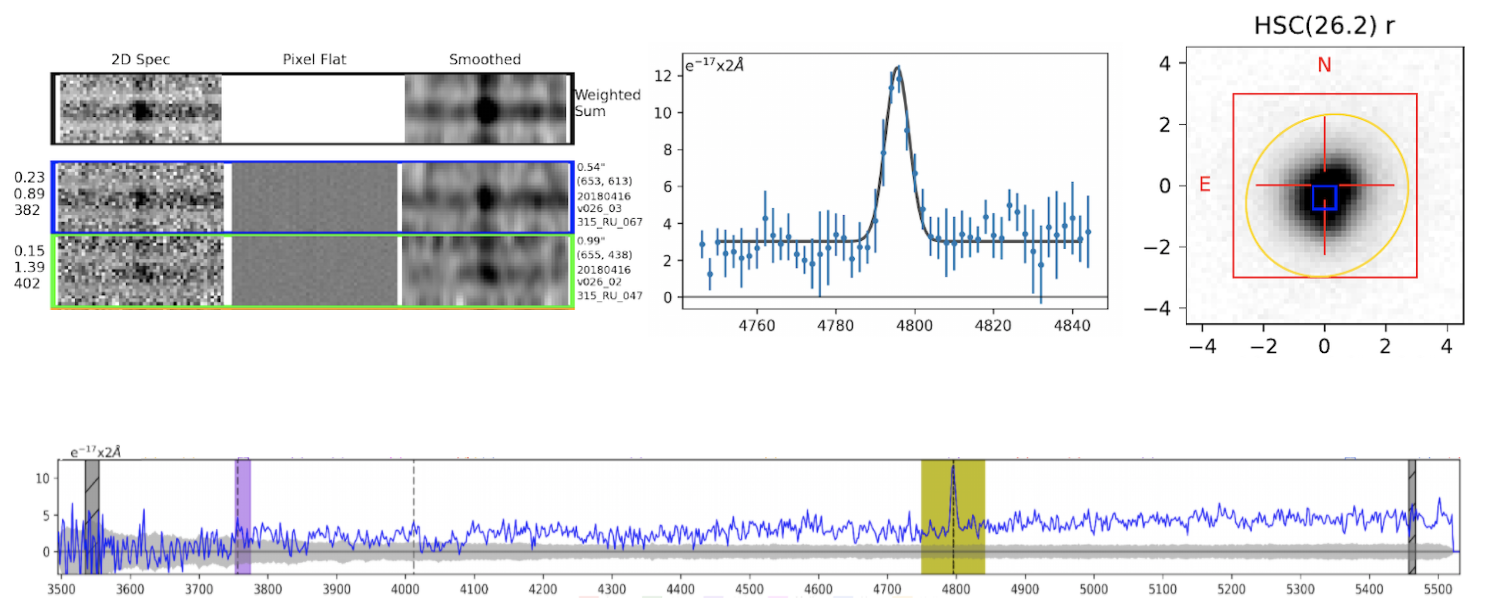}
    \caption{Returned HETDEX fibers, emission line fit, best available imaging of foreground source, and complete spectrum for the potential LAE lensed system rediscovered from \cite{Shu_2016}. The overall returned plots originate from ELiXer, which is discussed in detail in Section 3.2. As stated before, we are in agreement with the determined redshift found by \cite{Shu_2016}. \textit{Top Left}: Two best VIRUS fibers, pixel flats, smoothed VIRUS fibers, and weighted sum of fibers. The information on the sides of the fibers can be summarized as technical observation/pipeline information. The information on the right specifies the emission line on the CCD and the dither/shot information. The information on the left specifies the position of the fiber on the CCD and the fit to the expected fiber profile. \textit{Top Middle}: Gaussian fit to emission line with wavelength given in $\rm \AA$ and flux given in erg s$^{-1}$ cm$^{-2}$ over a $\rm 2 \AA$ window. \textit{Top Right}: Best available imaging at the location of lensed LAE candidate, which for this case is with the Hyper Suprime-Cam r band with the limited magnitude given. The panel distances are given in arcseconds. \textit{Bottom}: Full VIRUS spectrum of the potential lensed LAE system from \cite{Shu_2016}. Plotted is flux given in erg s$^{-1}$ cm$^{-2}$ over a $\rm 2 \AA$ bin versus wavelength given in $\rm \AA$. The yellow box with a dashed line denotes location of the identified potential Ly$\alpha$. The width of the yellow box is equal to width of the Gaussian cutout plot. The two gray columns represent skylines. The extra dashed lines and highlight colors indicate possible emission or absorption lines that pass a certain threshold from ELiXer. These are automatic in the ELiXer pipeline, but are not used for the analysis of this work as we are concerned with background LAE emission (see Section \ref{Extracting Spectra}). The gray shaded region in the 1-D spectra is a “rough” estimate of noise, not uncertainties on the flux measurement. The regions come from the standard deviation of flux per wavelength bin for a few hundred fibers spatially near the fibers of the detection (sigma clipping mostly excludes fibers that have significant signal from objects). Therefore the fibers utilized for the standard deviation are mostly (but not strictly) sky fibers.  The gray regions act only as a visual cue to the user, but is not otherwise used in any calculation. For faint emission lines, they can be very near this “noise”.}
    \label{fig:fig1}
\end{figure*}

We found a single BELLS GALLERY LAE candidate that fell on a HETDEX fiber, which is not surprising given the relatively small fraction of sky HETDEX has covered so far (compared to SDSS), the 1/4.5 filling factor of HETDEX IFUs (the IFUs, even after dithering to cover the inter-fiber gaps, only cover 1/4.5 of the HET focal plane; see Figure 3 of \cite{Hill_2016} and Figure 3 of \cite{Gebhardt_2021}, and the low surface density of strong gravitationally lensed systems. We
present the extracted HETDEX spectrum for this source
(SDSSJ144317.83+510721.0) in Figure \ref{fig:fig1}. From \cite{Shu_2016}, the redshift of the foreground source is $z = 0.5501$ and the redshift of the background source is $z=2.9443$ with a Ly$\alpha$ flux of $1.69 \pm 0.148 \times10^{-16}$ erg s$^{-1}$ cm$^{-2}$. \footnote{The reported error value was not present in the \cite{Shu_2016} publication, but obtained through private communication with the authors.}. From the extracted HETDEX spectrum, we found a background source redshift of $z = 2.9448$ with a Ly$\alpha$ flux of $3.60 \pm 0.40 \times 10^{-16}$ erg s$^{-1}$ cm$^{-2}$.

While our flux is higher,  a direct comparison between Ly$\alpha$ fluxes reported from \cite{Shu_2016} using BOSS and the current work using HETDEX is difficult. BOSS uses single fibers in a plug plate ($2\arcsec$) \citep{Drory_2015}, while HETDEX observations consist of 3-dither sequences with IFUs, thus collecting more of the total light. In particular, while a single fiber can be aperture corrected, this typically assumes a point-source profile, while the HETDEX IFU can be sensitive to more extended emission which has been observed in LAEs \citep[e.g,][]{Steidel_2011}, which is consistent with our $\sim$2$\times$ higher flux measurement. 

The utility of HETDEX for a new lensed search is due to the increased sensitivity of HETDEX. Specifically, although the lensed galaxies BELLS GALLERY finds are similar to what HETDEX can discover, the HETDEX spectroscopic sensitivity is greater, and is thus capable of detecting fainter emission lines down to $\sim 4\times10^{-17}$ erg s$^{-1}$ cm$^{-2}$ \citep{Adams_2011, Gebhardt_2021}, which corresponds roughly to 0.5L$^{\ast}_{Ly\alpha}$ at these redshifts \citep{De_La_Vieuville_2019}. For example, the faintest lensed LAE found in \cite{Lesser_2020}, an updated work built of off the original catalog of lensed LAEs from \cite{Shu_2016}, is $8.11\times10^{-17}$ erg s$^{-1}$ cm$^{-2}$ and was labeled with their lowest confidence interval. The factor of two greater sensitivity of HETDEX will allow the detection of intrinsically fainter Ly$\alpha$ emission, such that lensed LAEs should be more common with HETDEX completed, which we explore in the following sections.

\section{Building a Sample of Candidate Lensed LAEs} \label{Building a Sample of Candidate Lensed LAEs}

\subsection{Identifying Potential Foreground Lens Galaxies} \label{Identifying Potential Foreground Lens Galaxies}

To search for lensed LAEs in the HETDEX dataset, we elected to first identify a population of likely lenses and then explore these galaxies for emission lines. We began constructing a sample of foreground galaxies by obtaining all objects in DR16 from the Sloan Digital Sky Survey (SDSS) with clean photometry within a spectroscopic redshift of $0.4 \lesssim z \lesssim 0.7$. We elected to use SDSS to identify a population of likely lenses due to the sample size and continuum completion of SDSS (future versions of the HETDEX continuum catalog may make it possible to use HETDEX itself to identify these potential lensed systems). The redshift range was chosen by looking at prior LAE lensing work \citep{Bolton_2006, Brownstein_2012,Shu_2016, Lesser_2020}, in which all foreground lensing sources were common between $z \sim 0.4 - 0.7$; moreover, the availability for quality spectra, and thus quality statistics, is greater for the aforementioned redshift range compared with higher redshift foreground objects that would provide stronger gravitational lensing, but limit the overall sample. This redshift constraint yielded $\sim 1.4\times10^{6}$ galaxies.

Prior galaxy-galaxy lensed LAE studies \citep[e.g.,][]{Baldry_2004, Cao_2020} targeted red, non-star-forming galaxies for foreground lensing sources as they are prime environments to hunt for lensed LAEs. There are two central reasons in isolating red, non-starforming galaxies: 1) these galaxies tend to be more massive, and thus are more efficient at lensing and 2) the spectra of these galaxies typically contain few emission lines relative to star-forming galaxies, which simplifies the search for lensed emission lines overlaid in foreground spectra (see Section \ref{Identification Pipeline/Analysis}). The current work is acting as a pilot project for HETDEX, and so we are concerned with identifying potential strong lensed LAEs with red, non-star-forming galaxies as the lensing source, thus maximizing the likelihood of a true lensed LAE system through minimal foreground emission lines and strong lensing kernels. A more complete foreground sample, however, would include star-forming galaxies, although their inclusion increases the complexity of magnification estimates discussed in Section \ref{Magnification Determination} due to lensing modeling of spiral galaxies along with a greater entangling of foreground spectra with lensed emission lines. We do not search for potential lensed LAE systems behind star-forming galaxies in the current work.

\begin{figure}
    \centering
    \includegraphics[width=0.5\textwidth]{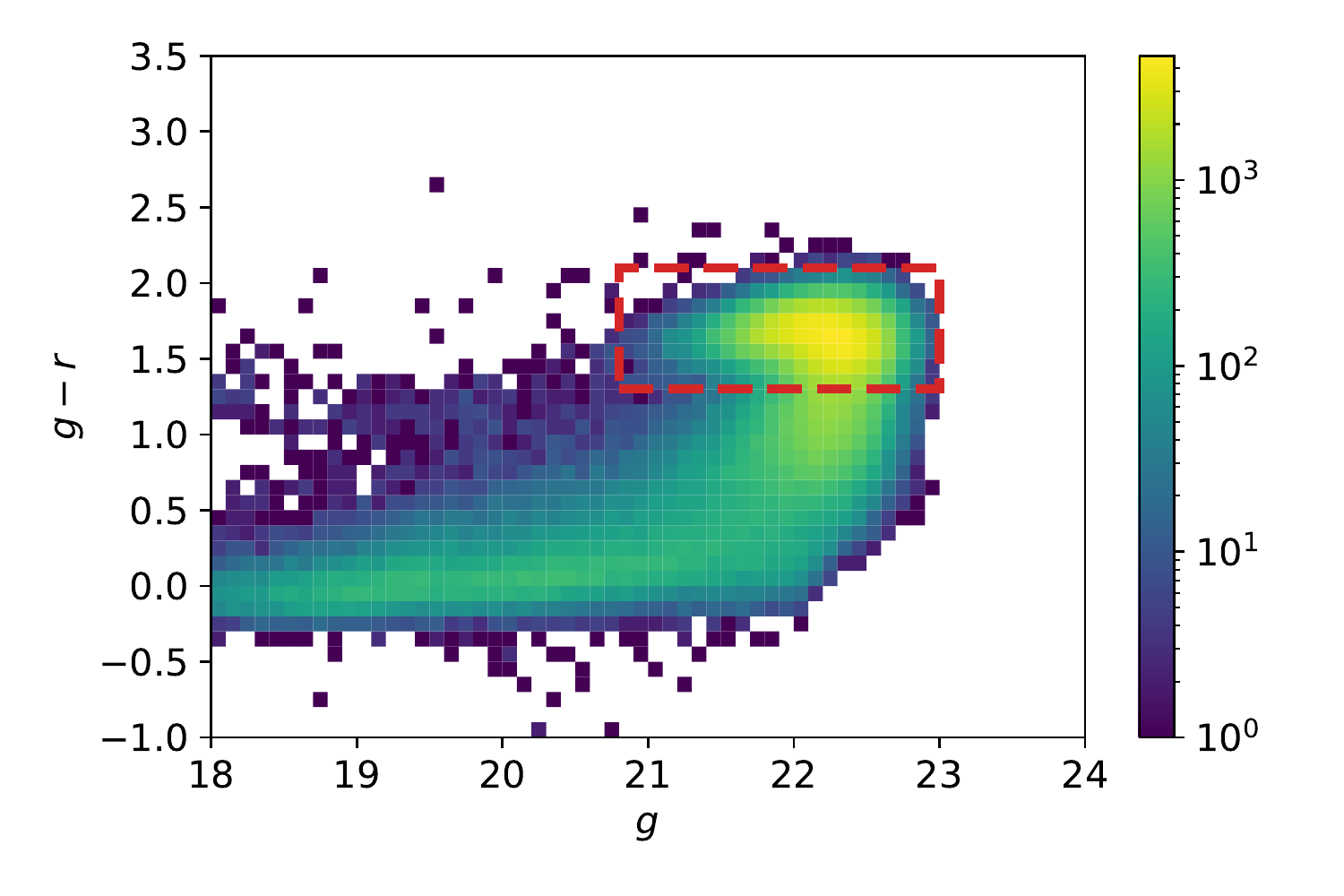}
    \caption{Color-Magnitude Diagram for all SDSS galaxies between the spectroscopic redshift range of $0.4 \lesssim z \lesssim 0.7$. The region between the red lines designates the foreground galaxy sample. The bottom, leftward tail of the CMD consists of predominately blue, star forming galaxies and QSO interlopers. The foreground galaxy sample region was chosen to eliminate star forming galaxies that could contaminate the sample. If any of our chosen potential lens systems exhibit emission lines from the lens galaxy, they were eliminated from the candidate catalog in the analysis stage, which is discussed in Section \ref{Identification Pipeline/Analysis}}
    \label{fig:fig2}
\end{figure}

To isolate red, non-star-forming galaxies from $\sim 1.4\times10^{6}$ redshift selected galaxies, we created a color-magnitude diagram (CMD) of the sample, which we present in Figure \ref{fig:fig2}. We applied color selections of $\rm 1.3 \leq g - r \leq 2.1$ and $\rm 20.8 \leq g \leq 23$ to isolate red, non-star-forming galaxies. The color-magnitude cuts were determined through visual inspection.  To select massive red galaxies, we defined the left side of the box to capture the high density tail end of red galaxies, the bottom side to separate the higher density region of bluer galaxies (directly below the red dashed line) and QSOs (leftward bottom tail) from the higher density region of red galaxies, and the upper side to minimize contaminants. From these color-magnitude cuts we isolated $\sim 130,000$ foreground galaxy candidates to search for within HETDEX coverage as described in Section \ref{Sec. 2}.

The selection of g-r vs. r would likely be a more appropriate choice here (and for future foreground selection procedures) as the Balmer/$\rm 4000 \AA$ break occurs over the chosen filters. However, g-r vs. g suffices in selecting massive galaxies and reducing contaminates. For any red galaxies with active star formation, any emission lines would be flagged through the procedure discussed in Section \ref{Identification Pipeline/Analysis}.

\subsection{Extracting Spectra} \label{Extracting Spectra}

With the foreground galaxy sample, the two potential problems from Section \ref{Searching for Known Lensed LAEs} - detections that did not coincide with the exact positions of foreground candidates and the positions of potential lensed LAEs falling between IFUs - persist, but were handled in a different method than what was stated in Section 2.2. Specifically, when we were proving that HETDEX is capable of detecting known potential lensed LAEs, we possessed the luxury of having the spectra and coordinates of established potential lensed LAEs, and thus were able to exploit the nearest VIRUS fiber to obtain corresponding spectra. With the foreground galaxy sample we designed, we no longer knew precisely where potential lensed LAEs resided, and thus were not able to immediately extract the result of a specific fiber. Therefore, we retained any foreground galaxy that was covered by HETDEX and employed a more robust method of obtaining spectra for each foreground candidate.

This more powerful, comprehensive method was the utilization of HETDEX’s Emission Line eXplorer \citep[ELiXer;][]{Davis_2021} software, which extracts spectra from the HETDEX HDR2 database. ELiXer is a data visualization, search, and diagnostic tool that combines HETDEX observational data and multiple photometric catalogs to provide a compact, user-friendly representation of emission line detections to assist the researcher in identifying line-emitting objects and their properties. ELiXer can be used on a desired sky position where it extracts an assortment of the nearest fibers and weights each fiber based on the distance from the given position and the PSF model, thus resulting in fiber coverage that can detect lensed Ly$\alpha$ emission rather than a single fiber that may or may not be covering the entire foreground galaxy.

In addition to extracting a spectrum centered on the position of the foreground lens galaxy, we used ELiXer to return any spectra within the HETDEX emission-line detection catalogue that resided within a $5\arcsec$ radius of the foreground galaxy's position. We established a search area around the foreground positions because exclusively searching at foreground positions limited the project to near perfectly aligned lenses. Extending the search area out to a $5\arcsec$ radius from the central foreground position allowed us to discover potential lensed LAEs in the strong, intermediate, or weak lensing regimes. Past $5\arcsec$, the galaxy-galaxy lensing kernel effectively diminishes to zero. Extending our search radius proved to be immensely fruitful and accounted for the majority of our final sample of potential lensed LAEs.

In addition to the added fiber coverage, ELiXer executes its own diagnostics to flag detections that may be spurious or non-astrophysical (bad pixels, meteors, cosmic rays, scattered light, etc) and returns line identification probabilities based on an aggregate weighting of various information sources (ratios of likelihoods based on equivalent widths, position and flux ratios from other possible emission lines, physical size with assumed redshifts, and more; see \cite{Leung_2017, Davis_2021, Farrow_2021, Gebhardt_2021} for more details).  The ELiXer success rate in identifying Ly$\alpha$ is $\sim 98 \%$ (\citep{Davis_2021, Gebhardt_2021}, though as we discuss in the following section we only use these probabities when considering potential lensed sources $>$2\arcsec\ from the lens galaxy (such that the lens continuum does not bias the classification).

\subsection{Identification Pipeline/Analysis} \label{Identification Pipeline/Analysis}

To discover potential lensed LAEs, we had to identify Ly$\alpha$ emission within spectra extracted on or close to the selected massive foreground galaxies. To do this, we identified emission lines in our sample by visual inspection. Specifically, out of the $\sim 130,000$ SDSS DR16 galaxies selected as lens candidates, $375$ were located within regions included in HETDEX HDR 2.1; this makes visual inspection of their extracted spectra feasible. We inspected ELiXer-based spectra extracted both at the position of the potential lens galaxy, as well as at the position of any nearby emission lines identified in the HDR2.1 emission line catalog (out to a $5\arcsec$ radius; see Section \ref{Extracting Spectra}).

\begin{figure*}
    \centering
    \includegraphics[width = \textwidth]{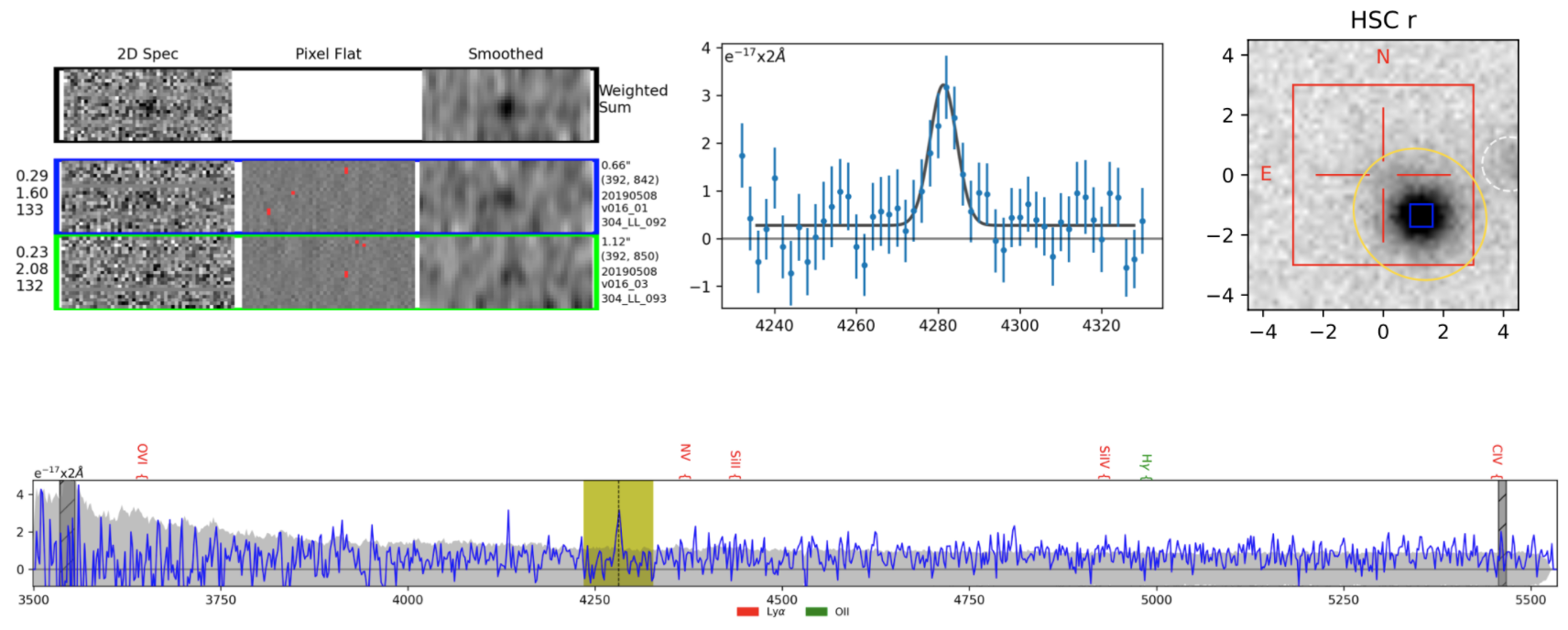}
    \caption{\textbf{Right Ascension: 167.706955 (ICRS, deg), Declination: 51.316486 (ICRS, deg)}. Same layout as described in Figure \ref{fig:fig1}. The red and green labels and the respective identifications above the 1D spectrum represent potential locations of emission/absorption lines if the identified line is either Ly$\alpha$ or [OII].  In this case no potential secondary line is seen, though this is nearly always the case when the primary line is identified as Ly$\alpha$. As described in Section \ref{Identification Pipeline/Analysis}, the spectroscopic redshift of the foreground galaxy is $z_{ls} = 0.4483$ with the identified emission line having $\rm \lambda = 4281.26 \AA$, which would make the identified emission line have a rest-frame wavelength of $2956.06 \rm \AA$. The closest neighboring rest-frame emission lines are HeI $\lambda 2945.11$ and OIII $\lambda 3132.79$.} 
    \label{fig:fig3}
\end{figure*}

Although our foreground galaxy sample of red, non-star forming galaxies was selected to have minimal star-forming emission lines, we could not immediately assume an identified emission line was from a lensed galaxy as emission lines can originate from processes unrelated to star-formation. Therefore, we determined the rest wavelength of each identified emission line by knowing the foreground galaxy’s spectroscopically determined redshift reported from SDSS (see Section \ref{SDSS Redshift Confirmation}). We then compared the rest wavelength against the full suite of strong and weak emission lines associated with galaxies, AGNs, and Quasi-Stellar Objects (QSOs) between $\rm \sim 700 \AA $ -- $ \rm 6000 \AA$ (e.g., [\ion{O}{3}]$\lambda 5007$, \ion{Ne}{5}$\lambda 3345.821$, \ion{Mg}{2}$\lambda 2802.705$) \footnote{The full list of rest wavelengths considered during this process can be found at \url{http://astronomy.nmsu.edu/drewski/tableofemissionlines.html}}. If the observed wavelength of the HETDEX emission line matched a known rest wavelength of an emission line at the foreground redshift, it was deemed to originate from the foreground galaxy, otherwise we concluded that the emission line was originating from a potentially lensed background galaxy. For example, in Figure \ref{fig:fig3} the wavelength of the identified emission line is at $\rm 4218 \AA$, which corresponds to a rest wavelength of $\rm 2912.4 \AA$ at the foreground redshift $\rm z = 0.4483$.  This does not correspond to any wavelengths from our adopted line list, thus likely pointing to a background lensed emission line. We performed this procedure on foreground sources and any neighboring HDR2.1 emission lines identified within $5\arcsec$ of the foreground galaxy's position.

\begin{figure*}
    \centering
    \includegraphics[width = \textwidth]{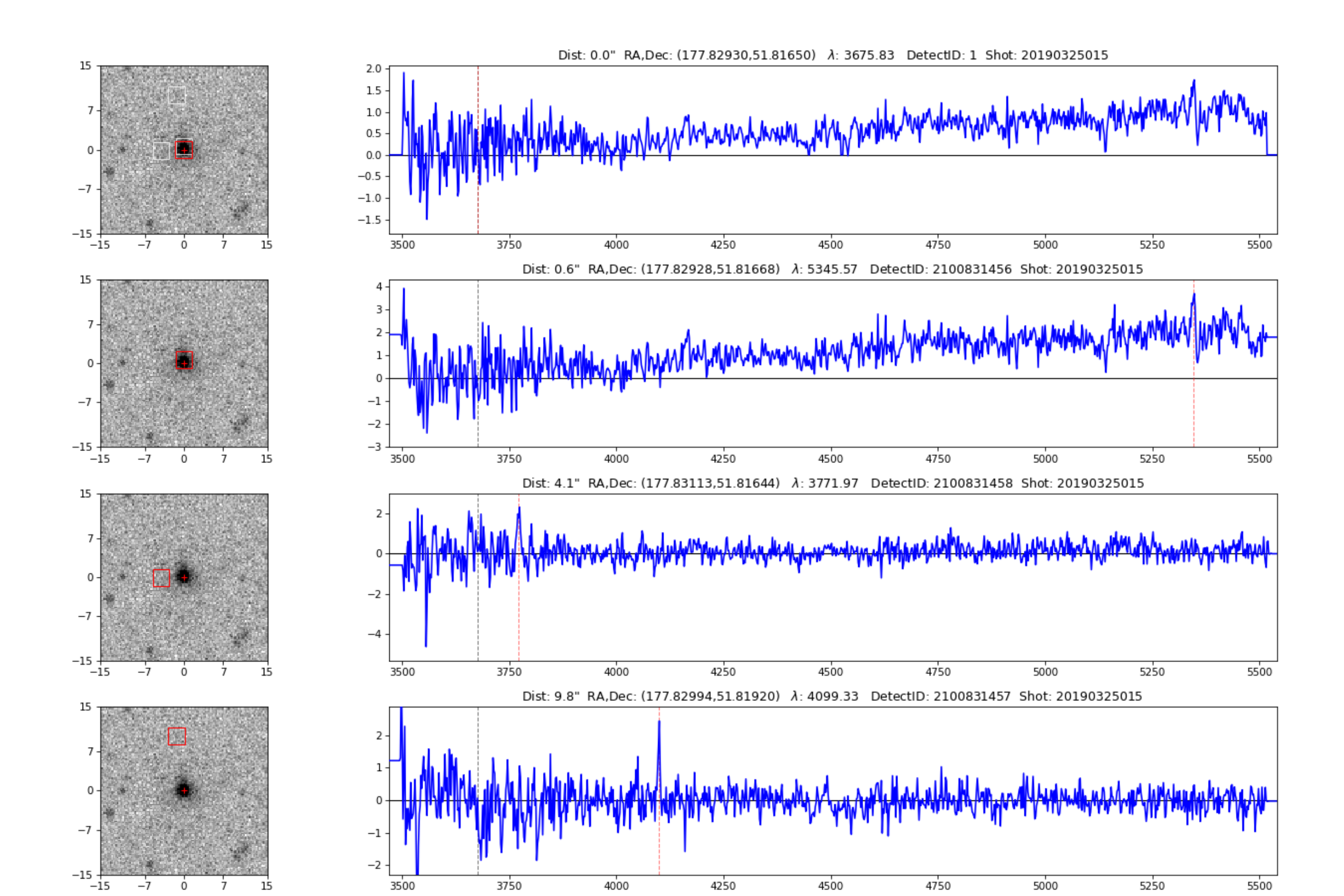}
    \caption{Example of neighborhood spectra returned using a $5\arcsec$ search radius from the foreground galaxy’s position using ELiXer for object $\rm 20190325v015-9900150001$. The first two spectra are oriented on the foreground galaxy and contain a non-detection (red line in top spectrum) and an [OII] emission (red line in second spectrum) respectively. The third spectrum is $4.2\arcsec$ away from the foreground galaxy and contains an emission line not present in the foreground galaxy spectrum and not originating from the foreground galaxy ($z_{lens} = 0.4340, \rm \lambda = 3772.96 \AA, \lambda_{0} = 2631.07 \AA$). We thus consider this detection a possible potential lensed LAE. The fourth spectrum, although containing an emission line, is $9.8\arcsec$ away from the foreground galaxy, and we thus do not consider it a viable potential lensed LAE. For each image panel the distances are given in arcseconds. For each spectra the flux is given in units of $\rm f_{\lambda} = 1\times10^{-17}$ erg s$^{-1}$ cm$^{-2}$. The gray dashed line at $\rm 3675.83 \rm \AA$ present in each spectrum represents ELiXer's first extraction of the region. In this case, the first extraction is a non-detection, but the wavelength of the first extraction is represented in the subsequent spectra. The red dashed line is the line for each detection corresponding to the red box in the associated imaging to the left. The neighbor search with ELiXer simply finds detections in the HETDEX detection catalog, so if a neighboring detection is on the same initial object you would expect to see the grey dashed line and the red line overlap (e.g., the top spectrum), or if an object has multiple detections the other detection would show up as a ``neighbor" (e.g., the second spectrum). }
    \label{fig:fig4}
\end{figure*}

For all emission lines, we required the emission line to have a $\geq$4.5$\sigma$ detection (measured by ELiXer), to attempt to rule out spurious sources.  We note this is similar to cuts used by other early HETDEX projects \citep[e.g.,][]{Indahl_2019}, and that the majority of our sources have signal-to-noise $>$5. We also visually removed any poor pixel flat subtractions (top left panel in Figure \ref{fig:fig3}), visually verified that the Gaussian fits (top middle panel in Figure \ref{fig:fig3}) were accurate, and visually checked for interlopers in the best available imaging (top right panel in Figure \ref{fig:fig3}). This process yielded nine potential lensed LAEs in the foreground-centered extractions, with an additional 17 lensed LAE candidates through the neighbor search method described. We present in Figure \ref{fig:fig4} an example of the neighbor search method.

The overwhelming majority of interlopers within our search were [OII]$\lambda\lambda3727,3729$ emitters, which is a common line seen in red galaxies in our foreground redshift search range \citep{Yan_2006, Lemaux_2010}. Although the rest wavelength determination step was imperative in uncovering potential lensed LAEs, it did not eliminate the possibility of discovering potential [OII] emitters or any other lensed emission lines. Unfortunately, at R $\sim 800$ (See Section \ref{HETDEX}), the Ly$\alpha$ asymmetry is rarely resolved, so we are not able to use line structure as a line identifier. Therefore, to ensure the candidate lensed emission lines were potential lensed Ly$\alpha$ emitters, we accessed the aforementioned probabilities calculated by ELiXer on the likelihood of the emission line being Ly$\alpha$. Specifically, ELiXer bases part of the analysis on \cite{Leung_2017} to calculate the Ly$\alpha$ probability, with a significant determining factor being the continuum brightness (and thus the line equivalent width).

If a lensed candidate was too close to the foreground galaxy then the probability of the emission being Ly$\alpha$ was skewed to a lower value due to the continuum of the foreground galaxy being included in the spectrum, biasing the line equivalent width. Knowing this, if a potential lensed candidate was within $\sim 2\arcsec$ of the foreground galaxy we did not immediately eliminate the candidate based on the ELiXer probabilities, but rather relied on our described identification of an emission line with no known associated rest wavelength.

The majority of the positions of our candidates, however, did not fall close enough to the foreground galaxy to skew the probabilities, and thus ELiXer probabilities were still reliable (candidates determined to have foreground continuum contamination are noted in Table \ref{Table 1}). Therefore, in addition to the emission lines needing a S/N $\geq 4.5$, we aimed for there to be a P(Ly$\alpha$)/P([OII]$\lambda\lambda3729,3727$) $\geq 3$ (a 3:1 ratio in favor of Ly$\alpha$) for emission lines to be considered confident lensed LAE candidates. This threshold of P(Ly$\alpha$)/P([OII]$\lambda\lambda3729,3727$) $\geq 3$ is explored in Davis et al.\ (in prep). In short, Davis et al. explore ELiXer's completeness of LAE identifications vs test data (i.e., recovery rate) under assumed conditions (e.g., P(Ly$\alpha$)/P([OII]$\lambda\lambda3729,3727$) $\geq 3$). ELiXer identifications are compared against $\sim 2500$ spec-z sources from various external catalogs, including $\sim 800$ sources from the Dark Energy Spectroscopic Instrument (DESI). They find a $\sim 95\%$ recovery rate for the P(Ly$\alpha$)/P([OII]$\lambda\lambda3729,3727$) $\geq 3$ condition assumption. Therefore, our P(Ly$\alpha$)/P([OII]$\lambda\lambda3729,3727$) $\geq 3$ threshold is a good lower limit to help narrow visual inspections and provide confidence to our identifications. Overall, however, the main evidence for potential lensed LAEs is the identification of an emission line inconsistent with the foreground redshift. 

We do note that Davis et al. (in prep) used a newer version of the HETDEX emission line catalog (HDR3) than we used (HDR2; HDR3 was not available at the time of our analysis).  However, broadly the data releases are the same and would not affect the current analysis. The differences lie in the available catalogs (e.g., greater areal coverage in HDR3) and some improvements in the automated ELiXer classifications, but the basic logic is the same. A repeat of the current work with HDR3 might obtain better (deeper) photometric imaging coverage, but the decision on whether an object is a background LAE candidate or not is unlikely to change.

In all, we discovered 26 potential lensed LAEs within the HETDEX data set. Nine LAEs fell on top or near the lens and another 17 were within $\sim 2^{\prime\prime}-5^{\prime\prime}$ of the foreground lensing galaxy. We present in Table \ref{Table 1} the details of each potential lensed LAE including a confidence level assigned by the authors based on the quality of the spectrum, S/N, P(LAE)/P(OII), and other returned ELiXer results to indicate which systems are most likely true lensed LAE systems. We present in Figures \ref{fig:fig3} and \ref{appendix fig 1}-\ref{appendix fig 3} the foreground spectrum, best available imaging, fiber weights, and the Gaussian fit to the Ly$\alpha$ emission line. The dashed lines in the spectra mark the location of the potential lensed Ly$\alpha$ emission. We present in Figures \ref{fig:fig5} and \ref{fig:fig6} the redshift distributions of both the foreground galaxies and the potential lensed LAEs.

\subsection{SDSS Redshift Confirmation} \label{SDSS Redshift Confirmation}

For the 26 potential lensed LAEs, we confirmed the reported SDSS spectroscopic foreground redshift through visual inspection to ensure the above analysis was valid. If the foreground redshifts were reported incorrectly then the emission lines we discovered could belong to the foreground galaxy instead of originating from a background, lensed LAE. We analyzed SDSS spectra of the 26 foreground galaxies by identifying emission and absorption lines (primarily Ca K ($\lambda3934.777$) and H ($\lambda3969.588$) absorption). In all, we found the redshifts to be consistent with SDSS redshifts, thus reinforcing the status of our reported systems as potential lensed LAEs.

Although we found the foreground SDSS redshifts to be correct, we discovered through the redshift vetting process that five of the foreground galaxies in the final lensed LAE candidate systems were identified by SDSS as QSOs. Therefore, unlike with ideal foreground candidates with a smaller number of expected emission lines, these QSO foreground galaxies could possess strong and/or less common emission lines. However, as mentioned in Section \ref{Identification Pipeline/Analysis}, we investigated all known strong and weak emission lines associated with galaxies, AGNs, and QSOs. Therefore, the emission lines identified we present are still likely to be associated with potential lensed LAEs. We present in Table \ref{table: Table 2} the QSOs in our sample, the rest wavelength of each identified emission line using the foreground redshift, nearby rest wavelength emission lines we vet, and the full width at half maximum (FWHM) of our identified emission lines. In addition to the rest wavelength of the identified emission line not matching any known rest wavelength, the identified emission lines in the QSO foreground spectra have a mean FWHM of $\rm <FWHM_{QSO}> = 6.95 \pm 1.26$, comparable to the average FWHM of our non-QSO sample of $\rm <FWHM_{nonQSO}> = 6.57 \pm 0.34$, thus further supporting the identified emission lines as potential lensed LAEs.

\begin{deluxetable}{cccc}

\tablecaption{QSOs in Foreground Sample}
\tablehead{HETDEX ID    &  $\lambda_{0}$ &  Nearest Lines & FWHM 
\\  & (\AA) & (\AA) & (\AA) }
\startdata
\\[0.05cm]
 20191029v020-9900140002 & 3087.3 & $\rm \substack{ \rm HeI \lambda2945.1,\\\rm OIII \lambda3132.8}$ & 4.5 $\pm$ 3.2\\[0.15cm]
 20190901v019-9900190001  &  2519.6 & $\rm \substack{ \rm CII] \lambda2324.7,\\\rm [FeXI] \lambda2648.7}$ & 11.1 $\pm$ 3.1\\[0.15cm]
 20180913v018-9900100003  &  2167.6 & $\rm \substack{ \rm NII] \lambda2142.8,\\\rm [OIII] \lambda2321.0}$ & 4.7 $\pm$ 3.3\\[0.15cm]
 20181205v014-9900080003  &  3090.5 & $\rm \substack{ \rm HeI \lambda2945.1,\\\rm OIII \lambda3132.8}$ & 9.7 $\pm$ 4.2\\[0.15cm]
 20190808v018-9900040003  &  2650.7 & $\rm \substack{ \rm [FeXI] \lambda2648.7,\\\rm HeII \lambda2733.3}$ & 4.8 $\pm$ 3.0\\
\enddata
\label{table: Table 2}
\tablecomments{This table lists the rest-frame wavelength at the lens redshift for the five lenses in our sample identified in SDSS as QSOs.  These rest-frame wavelengths do not correspond to any known QSO emission lines, making a high-redshift Ly$\alpha$ origin more likely.  In addition, none of these five emission lines have a FWHM resolved at $>$ 2 $\sigma$ significance (the FWHM of an unresolved line in HETDEX data is $\sim$5\AA). The five emission lines have a mean FWHM of $\rm <FWHM_{QSO}> = 6.95 \pm 1.26$, which is comparable to the average FWHM of our non-QSO sample of $\rm <FWHM_{nonQSO}> = 6.57 \pm 0.34$.}
\end{deluxetable}

\section{Magnification Determination} \label{Magnification Determination}

\subsection{Magnification} \label{Magnification}

Having a sample of lensed LAE candidates, we wanted to determine which lensing regimes our galaxies fell within to identify the most interesting candidates. Specifically, since we searched for lensed LAEs out to $5\arcsec$ from the foreground galaxies’ positions, we enabled the detection of strong, intermediate, and weak lensing of LAEs within our sample. To estimate the lensing magnification for our candidates, we followed \cite{Mason_2015}. Taking our lensing systems as Singular Isothermal Spheres (SIS), the magnification, $\mu$, is given by

\begin{equation} \label{mag eq}
    \mu = \frac{\lvert \theta \rvert}{\lvert \theta \rvert - \theta_{ER}},
\end{equation}
where $\theta_{ER}$ is the Einstein radius and $\theta$ is the image distance between the forefront galaxy and lensed source \cite{Mason_2015}. The SIS assumption ignores neighboring galaxies that potentially effect the magnification of the background source. For example, in Figure \ref{fig:fig7} there are systems where nearby galaxies are present that potentially modify the magnification strength; however, these galaxies are unlikely to play a significant role in magnification due their fainter fluxes, and thus likely lower masses relative to primary foreground galaxy. Ultimately, under the SIS assumption these proximate galaxies are ignored.

The $\mu$ determination yields what lensing regime each lensed LAE candidate falls within. Specifically, $\mu$ $>$ 2 is strong lensing, 1.4 $<$ $\mu$ $<$ 2 is intermediate lensing, and $\mu$ $<$ 1.4 is weak lensing.  We determined the Einstein radius from

 \begin{equation} \label{einstein radius}
     \theta_{ER}(\sigma, z) = 4\pi\frac{D_{ls}}{D_s}(\frac{\sigma}{c})^2, 
 \end{equation}
where $D_{ls}$ is the angular diameter distance between the lens and source, $D_s$ is the angular diameter distance between the observer to the source, $\sigma$ is the velocity dispersion of the lensed LAE candidate, and $c$ is the speed of light \cite{Mason_2015}. For $\sigma$, we utilized \cite{Mason_2015} correlation between velocity dispersion, redshift, and apparent magnitude of the foreground galaxies. We utilized the SDSS \textit{i'} band for the magnitude dependence in the velocity dispersion relation from Mason et al. (2015).

We accounted for the uncertainties in the \cite{Mason_2015} velocity dispersion$-$magnitude estimates and error induced by the PSF weighting of the VIRUS fibers ($\sim 0.5 \arcsec$; see \cite{Gebhardt_2021} for more details) by employing a Monte-Carlo error propagation technique. We evaluated Equation \ref{mag eq} $10,000$ times for each lensing candidate using values drawn randomly from normal distributions centered on the velocity dispersion error from \cite{Mason_2015}. Specifically, from \cite{Mason_2015}:

\begin{equation}
    \log(\sigma) = -0.1\times m + a\times \log (1+z) + b
\end{equation}
where $\sigma$ is the velocity dispersion, $m$ is the i-band magnitude, and a $\&$ b are constants. For
\begin{equation}
    z < 0.5: a = 2.26 \pm 0.79, b = 4.08 \pm 0.12 ,
\end{equation}
for
\begin{equation}
    0.5 < z < 1.0 : a = 0.93 \pm 0.13, b = 4.20 \pm 0.03 ,
\end{equation}
and for
\begin{equation}
   z > 1.0 : a = 1.02 \pm 0.15, b = 4.12 \pm 0.05 .
\end{equation}
The standard deviation of each $10,000$ runs per lensing candidate represents the $1\sigma$ error for the magnification determination and the median value represents $\mu$. For a comprehensive overview of the Monte-Carlo error propagation technique, please refer to \cite{andrae2010error}.

\begin{figure}
    \centering
    \includegraphics[width=0.5\textwidth]{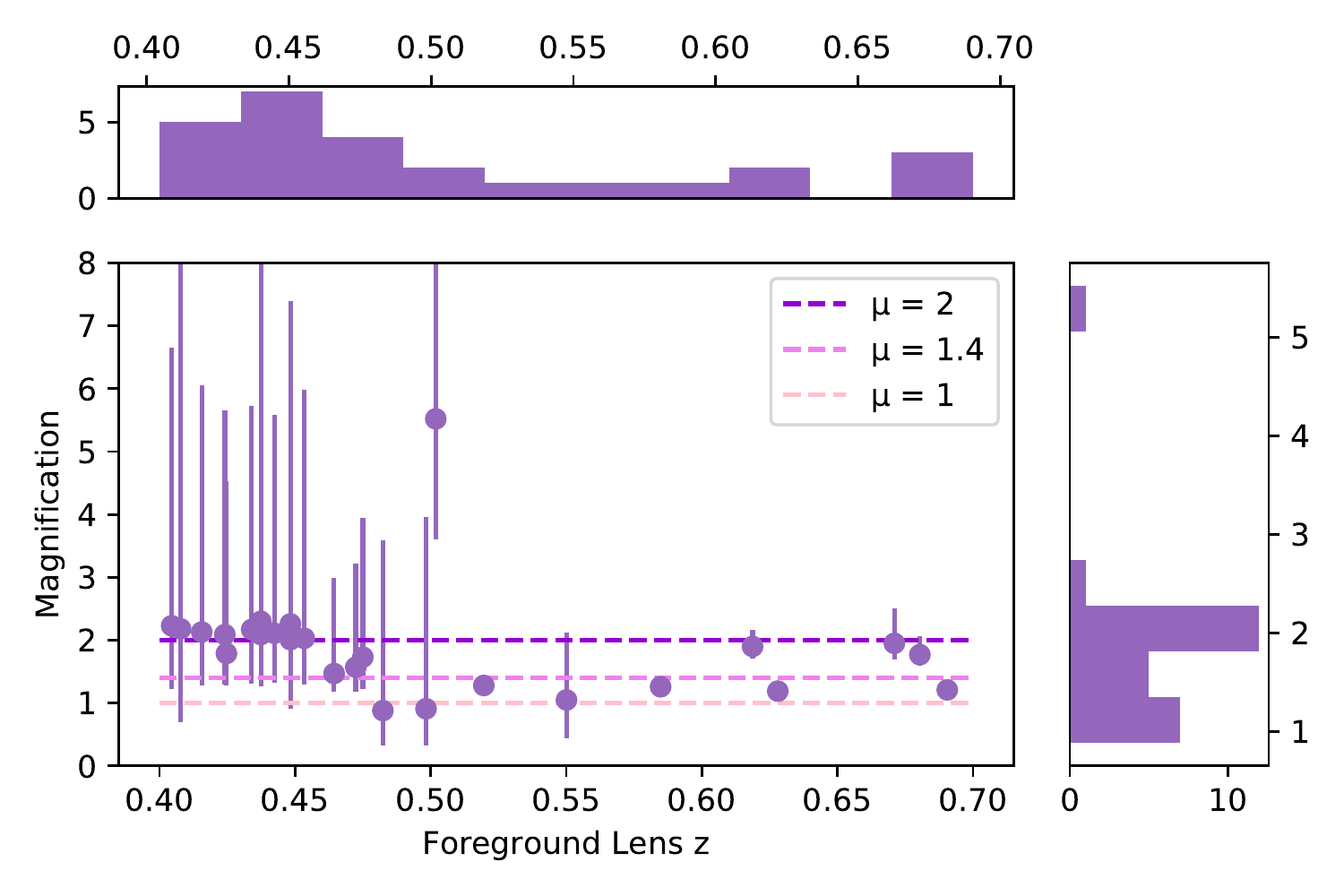}
    \caption{Distribution of foreground galaxy redshift and magnification. The three dotted lines represent the magnification cutoff for strong lensing, intermediate lensing, and weak lensing. As can be seen in the histogram to the right, a substantial sample of our candidates are in the strong lensing regime. The histogram presented above the scatter plot represents the foreground redshift distribution.}
    \label{fig:fig5}
\end{figure}

\begin{figure}
    \centering
    \includegraphics[width=0.5\textwidth]{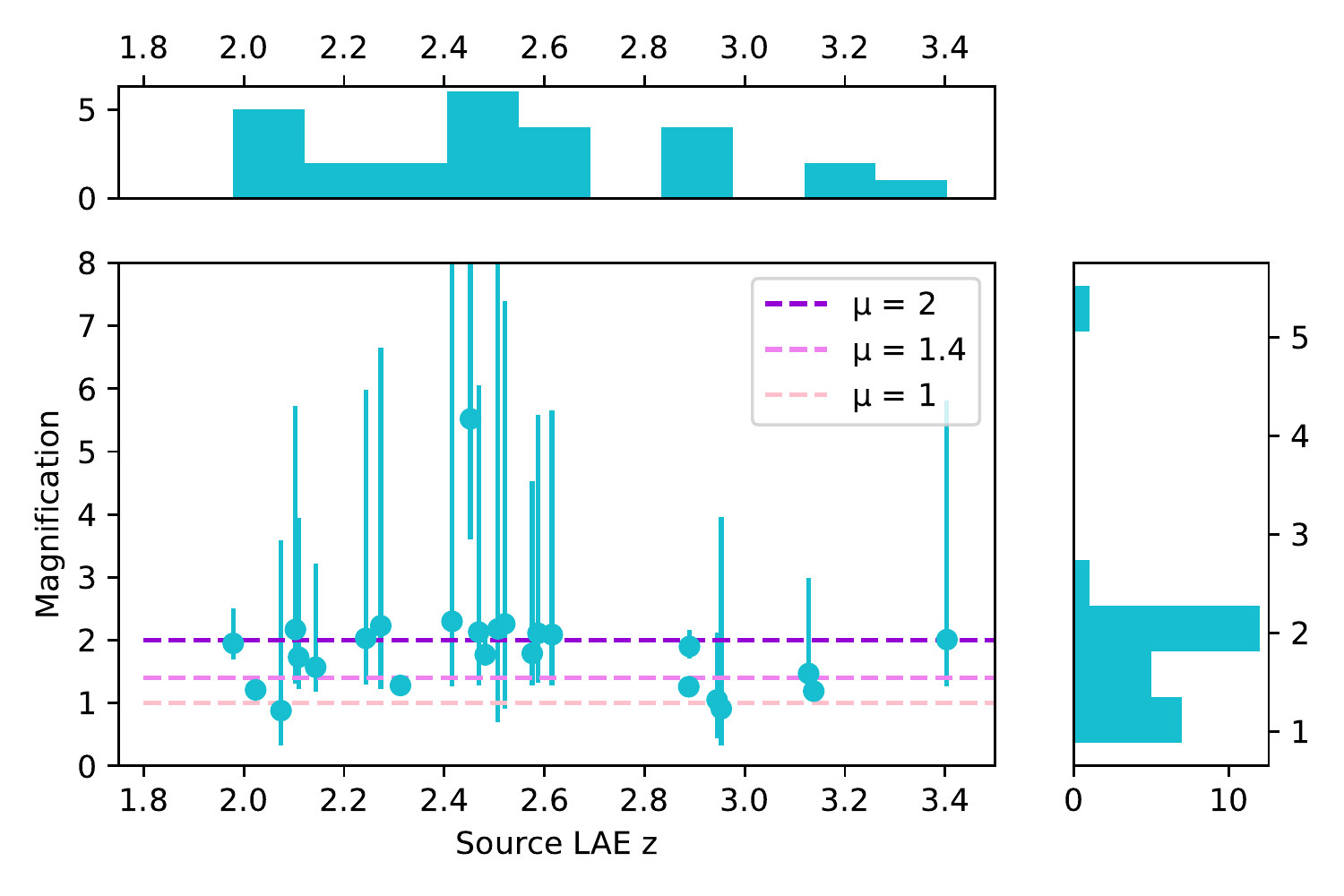}
    \caption{Distribution of source galaxy redshift and magnification with dotted lines and histograms representing the same info from Figure \ref{fig:fig5}.}
    \label{fig:fig6}
\end{figure}

\begin{figure*}[htb!]
\centering
\includegraphics[scale = 0.36]{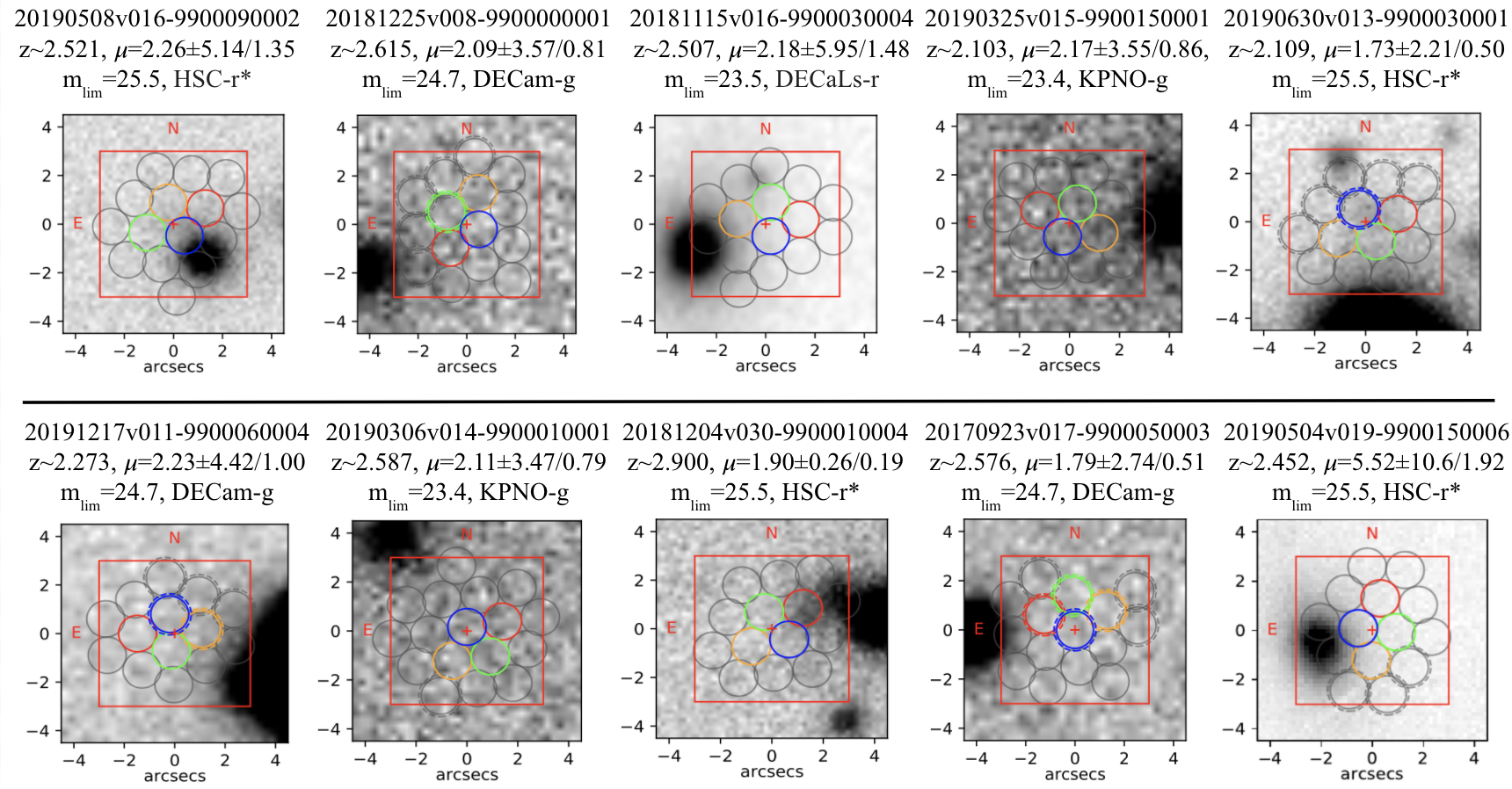}
\includegraphics[scale = 0.36]{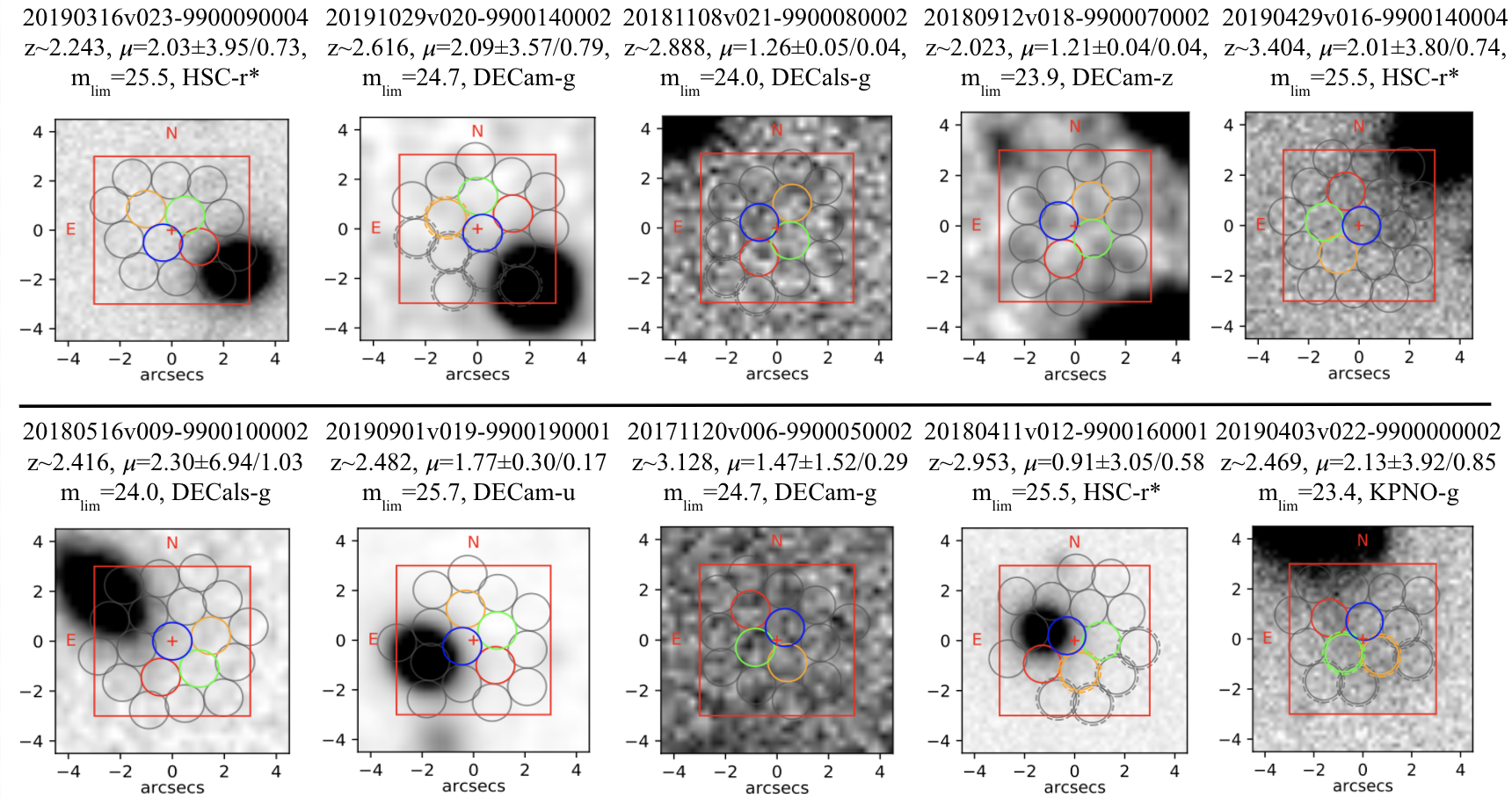}
\includegraphics[scale = 0.36]{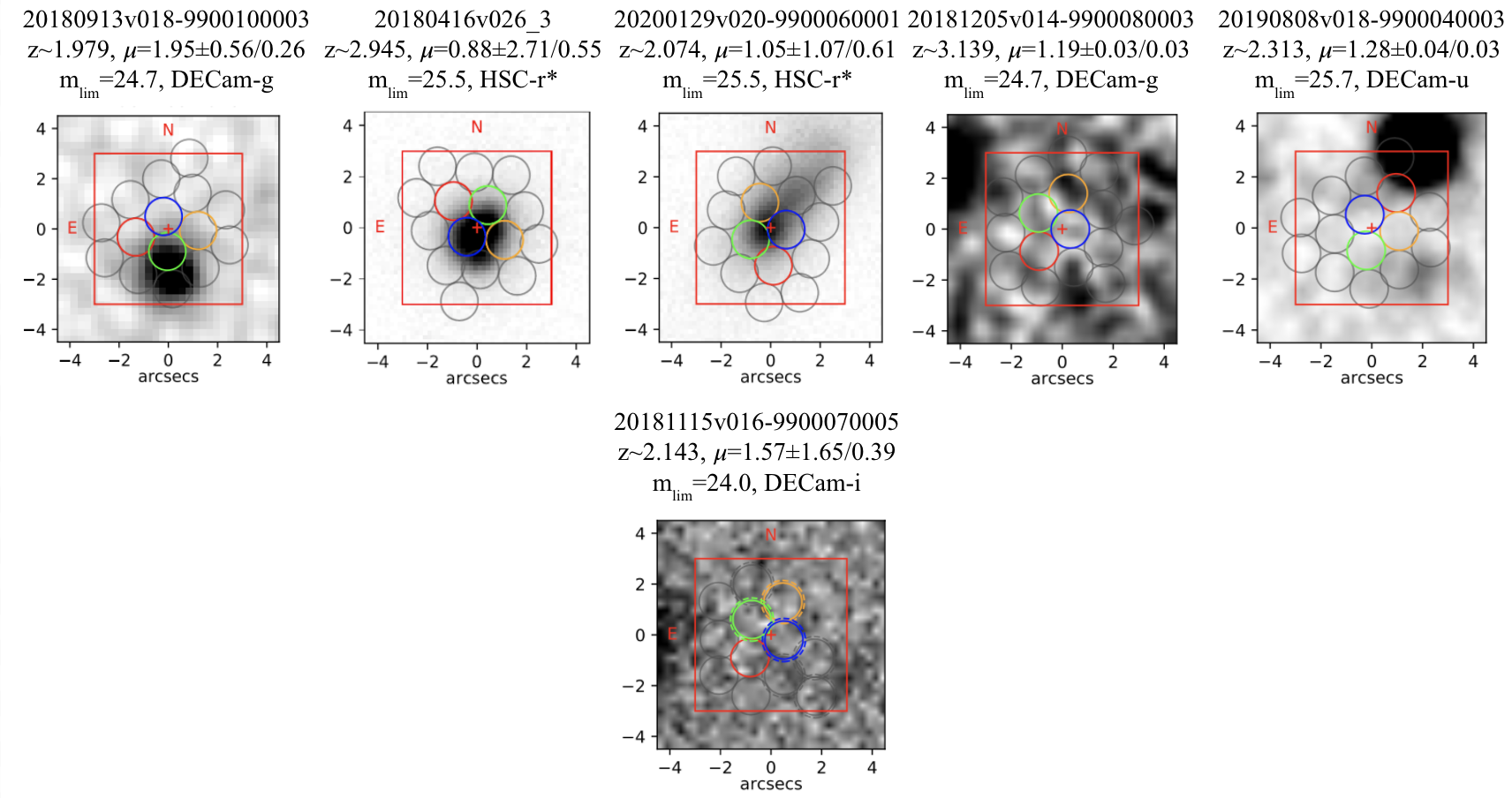}
    \caption{Best available imaging for the gravitationally lensed LAE candidates. We present the background lens redshift, magnification determined in Section \ref{Magnification Determination}, and the limiting magnitudes of each image, which are reported from the surveys themselves and are survey-wide averages (See Section \ref{HETDEX}). The cross-hair of each image focuses on the center of the background LAE candidate position, which is determined from the spectroscopic detection in the HETDEX spectral database found from ELiXer. The values along each axis represent arcseconds away from the central position of the image. The red box represents a $\pm 3\arcsec$ distance from the center of the image. The overlaid circles represent VIRUS fiber locations with the colors representing the fiber weights; strongest to weakest: blue, green, orange, and red. The imagining was taken directly from ELiXer, which has automatic contrast stretching, i.e., the bright foreground lens causes a poor stretch for identifying fainter regions. Moreover, due to the limited sensitivity of the best available imaging, the background LAE candidates being exceedingly faint, and the foreground object being bright, we do not expect a counterpart to the gravitationally lensed LAE location to be apparent in the current imaging. Nonetheless, the imaging is important to the current work as the magnification determination (Section \ref{Magnification Determination}) requires distances between the foreground lens and the background LAE candidate. We expect with deeper imaging for a counterpart to the Ly$\alpha$ detection to be discernible. The HSC-r imaging is custom to the HETDEX survey and is not associated with the Subaru Strategic Program.}
    \label{fig:fig7}
    
\end{figure*}

In total, we have $12$ candidate strong lensing systems, $7$ intermediate lensing systems, and $7$ weak lensing systems in our sample. Table \ref{Table 1} lists the magnification for each candidate, $z_{source}$, $z_{lens}$, S/N, positions, Ly$\alpha$ flux, LAE probability, and a confidence level, which was chosen by the authors based on the quality of spectra, S/N, and P(LAE)/P(OII). Figures \ref{fig:fig5} and \ref{fig:fig6} present the distribution of $\mu$ against corresponding redshift for foreground galaxies and background lensed LAE candidates respectively.

Follow up observations are needed to confirm the background galaxies as LAEs. Therefore, it was warranted to constrain the magnification of each system and establish a confidence level to note the lensing systems of most interest. Overall, the candidates with a high confidence level and with a magnification of $\rm \mu \geq 2$ are the best targets for follow up observations.

We present in Figure \ref{fig:fig7} the best available imaging, which we obtain using ELiXer, for the current potential lensed LAE catalog. Although the lensed LAEs are not obvious in the imaging, this is not unexpected due to the limiting magnitude of the current imaging, which is given for each candidate system in Figure \ref{fig:fig7}, being shallow relative to the expected magnitude of a faint, background LAE. Nonetheless, the available imaging aids in future studies spurring from the current catalog. Specifically, having the best available imaging, the magnification of each system, and the confidence level of each emission creates a solid foundation for future observations and work.

\subsection{Expected Number of LAEs}

In an effort to explore the validity of our sources as candidate lenses, we compare our observed number of 26 potential lenses to two predicted values. For both predictions, we compare their observed surface density to that in the overall internal HETDEX emission line catalog, which, to our chosen signal-to-noise
and Ly$\alpha$ probability cuts (S/N $\geq 4.5$ and P(LAE)/P(OII) $\geq 3$), has a surface density of $3.4$ LAEs/arcmin$^2$. 
We caution that this should be viewed as a lower limit for the expected true surface density of lensed sources, as the lensed sources are intrinsically fainter than un-lensed HETDEX sources, and thus will have higher intrinsic volume densities due to the shape of the Ly$\alpha$ luminosity function.
For our lensing search, we found 26 candidate lenses over an image-plane search area with a radius of $5\arcsec$ around 375 candidate lens systems with HETDEX spectra, resulting in a surface density of $\sim 3.2$ LAEs/arcmin$^2$. Thus, our search around potential lens systems yields approximately the expected source density, which is positive evidence against a significant spurious emission line fraction in our sample (which is expected, as we specifically targeted galaxies expected to have minimal emission line features).

As a second sanity check, we compare our observations to the predicted number of lensed galaxies. This requires us to \textit{a priori} assume the angular separation between the lens and the source in the source plane, denoted as $\rm \beta$. To decide on this distribution, we tested $\rm \beta$ distributions using the sample of 375 searched lens galaxies, in each case calculating the observed source-lens separation $\rm \theta$ using the known lens redshift, and a randomly drawn source lens redshift in the HETDEX redshift range of $1.9-3.5$. We found that assuming a flat random distribution of $\beta = 0-2\arcsec$ resulted in a $\theta$ distribution in agreement with that observed for our lensed galaxy sample, with a median of $\rm \theta \approx 4\arcsec$, and tails to both lower and higher values. A K-S test comparing these created distributions to our observed distributions gives a $\gtrsim50$\% probability that they were drawn from the same distribution.

Using this $\rm \beta$ distribution, we then ran 1000 Monte Carlo simulations. In each simulation, we considered whether a given potential lens galaxy would have a HETDEX LAE behind it, using the HETDEX surface density ($3.4$ LAEs/arcmin$^2$) and a search area with a radius of $3\arcsec$ (e.g., the maximum of our $\rm \beta$ distribution). This calculation implies that each lens system has a $2.6\%$ chance of having a true background LAE in the HETDEX redshift range. During each Monte Carlo simulation, we thus randomly assigned $2.6\%$ of the lens systems to have a background HETDEX LAE, and calculated the magnification. We then further calculated, for each simulation, the number of weak, intermediate, and strongly lensed systems. Using the median and standard deviation of these numbers across all 1000 simulations, we predict that we should have discovered $1 \pm 0.2$ weak, $1 \pm 1$ intermediate, and $7 \pm 2.7$ strongly lensed galaxies. Overall this expectation of 9 lensed galaxies is less than our sample size of 26 potential lenses, highlighting the need for followup deep and high-resolution imaging observations to confirm the lensed nature of these sources.

We note that our simple calculations are difficult for a variety of reasons, including the impact of subjective visual inspections during our candidate lensed galaxy selection process (where we implemented several qualitative cuts to arrive at a robust list of emission lines which are not encapsulated in the internal HETDEX emission line catalog surface density), the impact of lensing distortion of the search volume (for the image-plane prediction), and also that the area within our search radius to where we are sensitive to emission lines is likely less than the full area due to confusion with the light (and associated noise) from the lens galaxy itself. We conclude that followup observations of our candidate lens systems, ideally high-resolution imaging with, e.g., the {\it Hubble Space Telescope}, are needed to confirm the validity of these systems.

\begin{deluxetable*}{cccccccccc}
\tabletypesize{\footnotesize}
\tablewidth{0pt}
\tablecaption{Gravitationally Lensed LAEs Candidates \label{tab:sample}}
\tablehead{\colhead{HETDEX ID} & \colhead{Right Ascension} & \colhead{Declination} & \colhead{$z_{source}$}& \colhead{$z_{lens}$} & \colhead{$f_{Ly\alpha}$} & \colhead{S/N} & \colhead{$\mu^{a}$} & \colhead{$\frac{P(LAE)}{P(OII)}^{b}$} & \colhead{Confidence Level$^{c}$} \\
\colhead{} & \colhead{(J2000.0)} & \colhead{(J2000.0)} & \colhead{}& \colhead{} & \colhead{(erg s$^{-1}$ cm$^{-2}$)} & \colhead{} & \colhead{} & \colhead{} & \colhead{}}
\startdata
20190508v016-9900090002 & 167.706955 & 51.316486 & 2.5208 & 0.4483 &1.20($\pm$0.29)$\times 10^{-16}$ & 10.2 & 2.26 $\substack{+5.14 \\ -1.35}$\ & 2.196$^d$ & 5/5\\
20181225v008-9900000001 & $\phantom{0}$29.659389 & 0.416671 & 2.6151 & 0.4374 & 8.00($\pm$2.10)$\times 10^{-17}$ & 6.4 & 2.09 $\substack{+3.57 \\ -0.81}$\ & 20.29 & 5/5\\
20181115v016-9900030004 & $\phantom{0}$36.339851 & 0.021793 & 2.5073 & 0.4078 & 1.10($\pm$0.29)$\times 10^{-16}$ & 5.6 & 2.18 $\substack{+5.95 \\ -1.48}$\ & 7.971 & 5/5\\
20190325v015-9900150001 & 177.831100 & 51.816456 & 2.1028 & 0.4340 &1.30($\pm$0.32)$\times 10^{-16}$ & 4.3$^e$ &  2.17 $\substack{+3.55 \\ -0.86}$\ & 1000$^f$ & 4/5\\
20190630v013-9900030001 & 203.432022 & 51.853287 & 2.1091 & 0.4750 &1.60($\pm$0.36)$\times 10^{-16}$ & 5.7 & 1.73 $\substack{+2.21 \\ -0.50}$\ & 1000$^f$ & 5/5\\
20191217v011-9900060004 & $\phantom{0}$26.690495 & 0.025604 & 2.2732 & 0.4045 &3.20($\pm$0.85)$\times 10^{-16}$ & 6.7 & 2.23 $\substack{+4.42 \\ -1.00}$\ & 14.34 & 5/5\\
20190306v014-9900010001 & 165.313126 & 51.828682 & 2.5871 & 0.4424 & 2.50($\pm$0.44)$\times 10^{-16}$ & 6.1 & 2.11 $\substack{+3.47 \\ -0.79}$\ & 1000$^f$ & 5/5\\
20181204v030-9900010004 & 174.282944 & 51.645485 & 2.8997 & 0.6187 & 7.60($\pm$1.90)$\times 10^{-17}$ & 7.0 & 1.90 $\substack{+0.26 \\ -0.19}$\ & 21.34 & 5/5\\
20170923v017-9900050003 & $\phantom{0}$27.731714 & $-$0.085593 & 2.5757 & 0.4247 & 1.20($\pm$0.30)$\times 10^{-16}$ & 7.2 & 1.79 $\substack{+2.74 \\ -0.51}$\ & 506.4 & 5/5\\
20190504v019-9900150006 & 166.138824 & 51.107941 & 2.4523 & 0.5019 & 4.50($\pm$1.60)$\times 10^{-17}$ & 5.8 & 5.52 $\substack{+10.6 \\ -1.92}$\ & 0.954$^{d}$ & 3/5\\
20190316v023-9900090004 & 203.483170 & 51.590237 & 2.2432 & 0.4534 & 2.10($\pm$0.52)$\times 10^{-16}$ & 5.1 & 2.03 $\substack{+3.95 \\ -0.73}$\ & 1.222$^{d}$ & 3/5\\
20191029v020-9900140002$^g$ & $\phantom{0}$33.200455 & $-$0.381712 & 2.6156 & 0.4241 & 7.70($\pm$2.10)$\times 10^{-17}$ & 4.5 & 2.09 $\substack{+3.57 \\ -0.79}$\ & 0.219$^{d}$ & 3/5\\
20181108v021-9900080002 & $\phantom{0}$36.746841 & $-$0.094904 & 2.8882 & 0.5848 & 5.70($\pm$1.90)$\times 10^{-17}$ & 4.9 & 1.26 $\substack{+0.05 \\ -0.04}$\ & 8.524 & 4/5\\
20180912v018-9900070002 & $\phantom{0}$15.943008 & $-$0.074252 & 2.0233 & 0.6905 & 2.20($\pm$0.68)$\times 10^{-16}$ & 6.3 & 1.21 $\substack{+0.04 \\ -0.04}$\ & 1000$^f$ & 5/5\\
20190429v016-9900140004 & 165.877670 & 51.447983 & 3.4038 & 0.4483 & 2.00($\pm$0.55)$\times 10^{-16}$ & 5.9 & 2.01 $\substack{+3.80 \\ -0.74}$\ & 9.656 & 4/5\\
20180516v009-9900100002 & 230.091782 & 51.188099 & 2.4155 & 0.4374 & 1.00($\pm$0.31)$\times 10^{-16}$ & 5.8 & 2.30 $\substack{+6.94 \\ -1.03}$\ & 14.13 & 4/5\\
20190901v019-9900190001$^g$ & $\phantom{0}$14.839665 & $-$0.503009 & 2.4819 & 0.6804 & 1.30($\pm$0.36)$\times 10^{-16}$ & 5.5 & 1.77 $\substack{+0.30 \\ -0.17}$\ & 0.392$^{d}$ & 3/5\\
20171120v006-9900050002 & $\phantom{0}$28.657616 & 0.058895 & 3.1276 & 0.4644 & 8.00($\pm$2.80)$\times 10^{-17}$ & 5.0 & 1.47 $\substack{+1.52 \\ -0.29}$\ & 9.911 & 3/5\\
20180411v012-9900160001 & 159.745132 & 51.019459 & 2.9531 & 0.4983 & 7.70($\pm$2.60)$\times 10^{-17}$ & 5.0 & 0.91 $\substack{+3.05 \\ -0.58}$\ & 0.109$^{d}$ & 2/5\\
20190403v022-9900000002 & 194.097626 & 51.888298 & 2.4688 & 0.4156 & 5.90($\pm$1.70)$\times 10^{-17}$ & 5.6 & 2.13 $\substack{+3.92 \\ -0.85}$\ & 124.9 & 3/5\\
20180913v018-9900100003$^g$ & $\phantom{0}$33.088234 & $-$0.156773 & 1.9786 & 0.6710 & 1.10($\pm$0.45)$\times 10^{-16}$ & 5.1 & 1.95 $\substack{+0.56 \\ -0.26}$\ & 1000$^f$ & 3/5\\
20200129v020-9900060001 & 164.699598 & 50.323509 & 2.0740 & 0.4824 & 8.10($\pm$4.40)$\times 10^{-17}$ & 4.1$^{e}$ &  0.88 $\substack{+2.71 \\ -0.55}$\ & 1000$^f$ & 2/5\\
20180416v026-3 & $\phantom{0}$220.824325 & 51.122505 & 2.9448 & 0.5501 & 3.60($\pm$0.40)$\times 10^{-16}$ & 10.4 & 1.05 $\substack{+1.07 \\ -0.61}$\ & 0.002$^d$ & 3/5\\
20181205v014-9900080003$^g$ & $\phantom{0}$35.701439 & 0.414818 & 3.1378 & 0.6280 & 1.90($\pm$0.43)$\times 10^{-16}$ & 6.9 & 1.19 $\substack{+0.03 \\ -0.03}$\ & 1000$^f$ & 3/5\\
20190808v018-9900040003$^g$ & $\phantom{0}$ 17.096273 & 0.590635 & 2.3125 & 0.5196 & 5.30($\pm$2.70)$\times 10^{-17}$ & 4.8 & 1.28 $\substack{+0.04 \\ -0.03}$\ & 1000$^f$ & 3/5\\
20181115v016-9900070005 & $\phantom{0}$ 36.528481 & -0.033337 & 2.1432 & 0.4724 & 2.10($\pm1.30$)$\times$ $10^{-16}$ & 5.0 & 1.57 $\substack{+1.65 \\ -0.39}$ \ & $1000^f$ & 2/5 
\enddata
\vspace{0.5cm}
\tablecomments{$^a$ Magnification Determination determined from section 6. $^b$ Returned ELiXer probabilities. $^c$ Confidence level assigned to each candidate based on visual inspection of spectrum, S/N, P(LAE)/P(OII), and other returned ELiXer results. $^d$ Example where P(LAE)/P(OII) is $<$ 3 due to LAE candidate residing too close to the forefront galaxy, thus skewing the probability calculation. $^e$ Candidate where the S/N is beneath the catalog cutoff; however, as can be seen in Figure \ref{appendix fig 2}, the emission is clear. The lower S/N can be attributed to the emission residing in the blue region of the spectrum where noise drastically rises. $^f$ Returned probability ratio maximum from ELiXer. $^g$ Foreground QSOs.}
\label{Table 1}
\end{deluxetable*}

\section{Summary/Future} \label{Summary}

We have discovered 26 potential gravitationally lensed LAEs within the HETDEX data set. We isolated these potential lensing systems by visually inspecting fiber weighted spectra extracted with ELiXer and utilizing emission line probabilities determined from ELiXer. In addition to finding the potential lensed LAE systems, we determined the magnification for each potential lensed LAE system. We found $12$ potential LAEs within the strong lensing regime, $7$ within the intermediate regime, and $7$ within the weak lensing regime. Based on the visual inspection of the spectra, magnification, emission line probabilities, S/N, and emission line fits, we assigned a confidence level of LAE likelihood to each candidate, which we present in Table \ref{Table 1}.

HETDEX is only $\sim 35\%$ complete. Therefore, as HETDEX progresses, so will the number of potential gravitationally lensed LAEs observed by HETDEX. For this project, we set out not only to provide a list of potential gravitationally lensed LAEs, but also to establish a foundation for an analogous project to be launched once HETDEX is complete. 

The lensed LAE candidates presented in this paper require observational followup to confirm the lensing status. If confirmed, the science that can be performed on the background LAEs is wide ranging. For research concerned with the epoch of reionization, the added observational detail and number of lensed LAEs enables stronger constraints on the relative contribution of star-forming galaxies to cosmic reionization. For research concerned with galaxy evolution, observing lensed LAEs that would be too faint to observe if they had not been gravitationally lensed enables improvements on luminosity functions such as constraining the faint-end slope, which remains unconstrained. Moreover, lensed LAEs give a more detailed image of Milky Way progenitors. For example, \cite{Shu_2016} used strongly lensed LAEs to obtain the morphology of the LAEs down to $\sim 100$ pc regions. In addition to the morphology of LAEs, the line profiles of Ly$\alpha$ emission are still not fully understood \citep{Yamada_2012}. Specifically, the origin of Ly$\alpha$ emission does not have a dominant identified source and the process of Ly$\alpha$ photon escape from the galaxies is labyrinthine \citep{Yamada_2012}, thus leading to varying Ly$\alpha$ profiles. Possessing detailed observations on the candidates identified in this paper can provide insight for the diverse theoretical models of Ly$\alpha$ escape and sources.

In all, the work presented here lays the foundation for subsequent work to be performed once HETDEX is completed, identifies potential gravitationally lensed LAE systems, and determines the magnification for each lensing system.

\section{Acknowledgments}

We thank Sangeeta Malhotra for stimulating conversations which led to this work, and the staff at McDonald Observatory for their tremendous effort to build this new VIRUS instrument. HETDEX is led by the University of Texas at Austin McDonald Observatory and Department of Astronomy with participation from the Max-Planck-Institut f$\rm \Ddot{u}$r Extraterrestriche Physik (MPE), Leibniz-Institut f$\rm \Ddot{u}$e Astrophysik Potsdam (AIP), Texas A$\&$M University, Pennsylvania State University, Institut f$\rm \Ddot{u}$r Astrophysik G$\rm \Ddot{o}$ttingen, The University of Oxford, Max-Planck-Institut f$\rm \Ddot{u}$r Astrophysik (MPA), The University of Tokyo and Missouri University of Science and Technology. In addition to Institutional support, HETDEX is funded by the National Science Foundation (grant AST-0926815), the State of Texas, the US Air Force (AFRL FA9451-04-2- 0355), and generous support from private individuals and foundations. The observations were obtained with the Hobby-Eberly Telescope (HET), which is a joint project of the University of Texas at Austin, the Pennsylvania State University, Ludwig-Maximilians-Universit$\rm \Ddot{a}t$ M$\rm \Ddot{u}$nchen, and Georg-August-Universit$\rm \Ddot{a}$t G$\rm \Ddot{0}$ttingen. The HET is named in honor of its principal benefactors, William P. Hobby and Robert E. Eberly. The authors acknowledge the Texas Advanced Computing Center (TACC) at The University of Texas at Austin for providing high performance computing, visualization, and storage resources that have contributed to the research results reported within this paper. URL: http://www.tacc.utexas.edu
The Institute for Gravitation and the Cosmos is supported by the Eberly College of Science and the Office of the Senior Vice President for Research at the Pennsylvania State University.
We generously thank the University of Texas, the HETDEX collaboration, the University of Texas High Z Group, and The John W. Cox Endowment for Advanced Studies in Astronomy. Isaac Laseter and Steven Finkelstein acknowledge support from the National Science Foundation, through grant AST-1908817.

\newpage
\clearpage
\bibliographystyle{aasjournal}
\bibliography{bibliography.bib}


\appendix

\begin{figure*}[h]
    \centering
\includegraphics[ width=\textwidth]{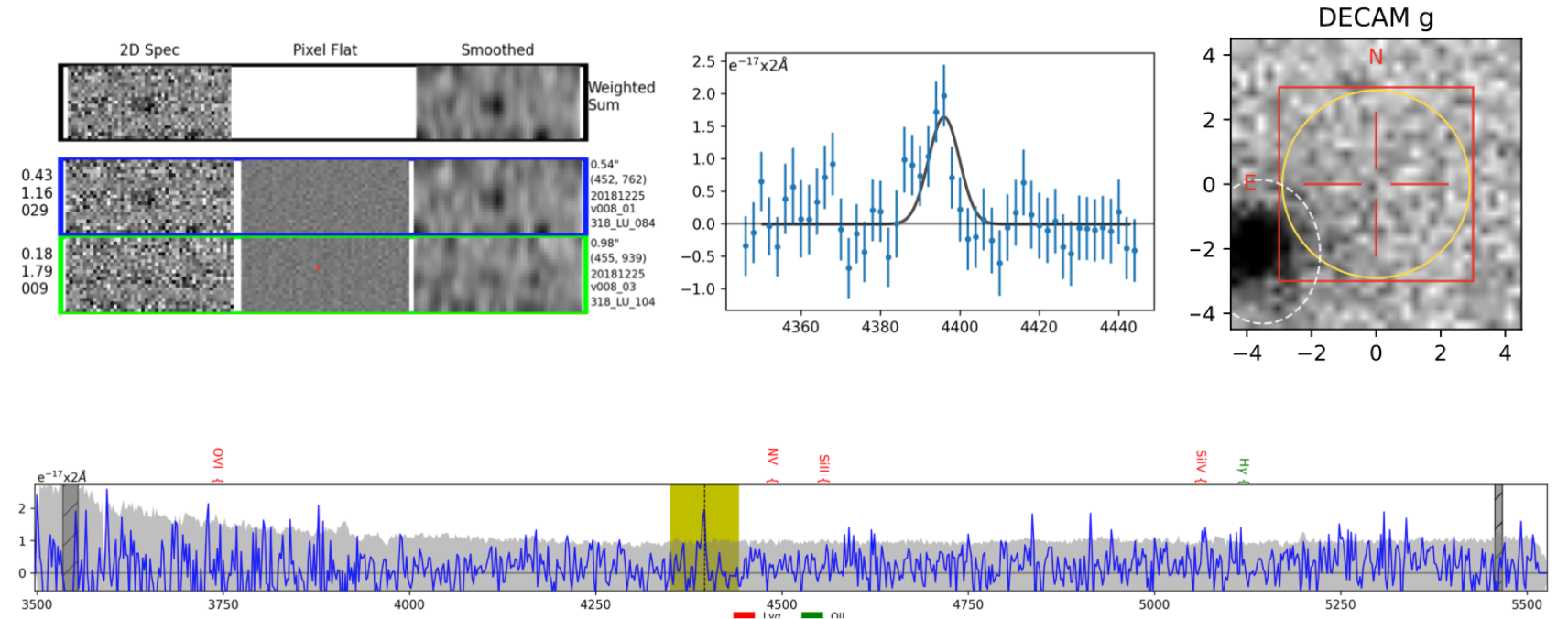}
    \caption{\textbf{20181225v008-9900000001.} \emph{Top Left}: Two best VIRUS fibers, pixel flats, smoothed VIRUS fibers, and weighted sum of fibers. \emph{Top Middle}: Gaussian fit to emission line. \emph{Top Right}: Best available imaging on location of lensed LAE candidate. \emph{Bottom}: Full spectrum of lensed LAE candidate. The yellow box with a dashed line denotes location of Ly$\alpha$.}
    \label{appendix fig 1}
\end{figure*}

\begin{figure*}
    \centering
\includegraphics[ width=\textwidth]{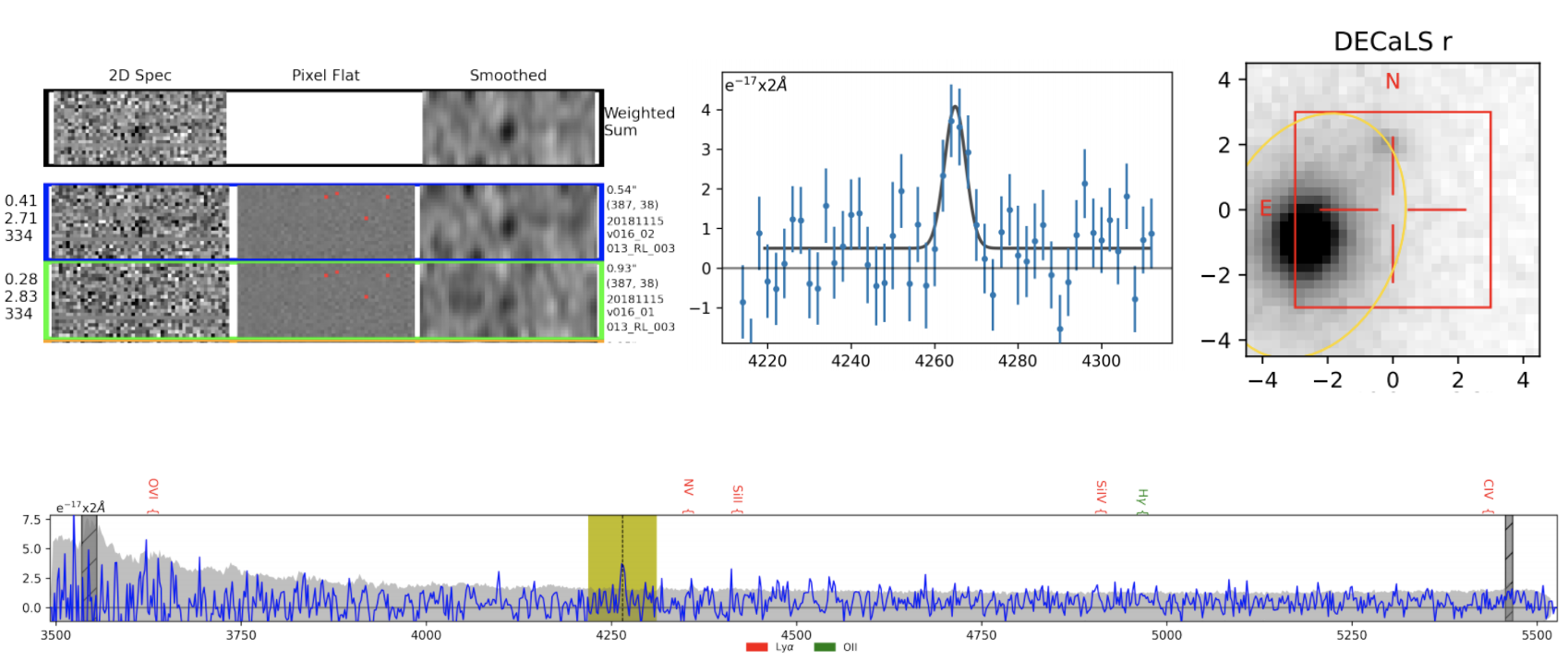}
    \caption{\textbf{20181115v016-9900030004.} \emph{Top Left}: Two best VIRUS fibers, pixel flats, smoothed VIRUS fibers, and weighted sum of fibers. \emph{Top Middle}: Gaussian fit to emission line. \emph{Top Right}: Best available imaging on location of lensed LAE candidate. \emph{Bottom}: Full spectrum of lensed LAE candidate. The yellow box with a dashed line denotes location of Ly$\alpha$.}
\end{figure*}

\begin{figure*}
    \centering
\includegraphics[ width=\textwidth]{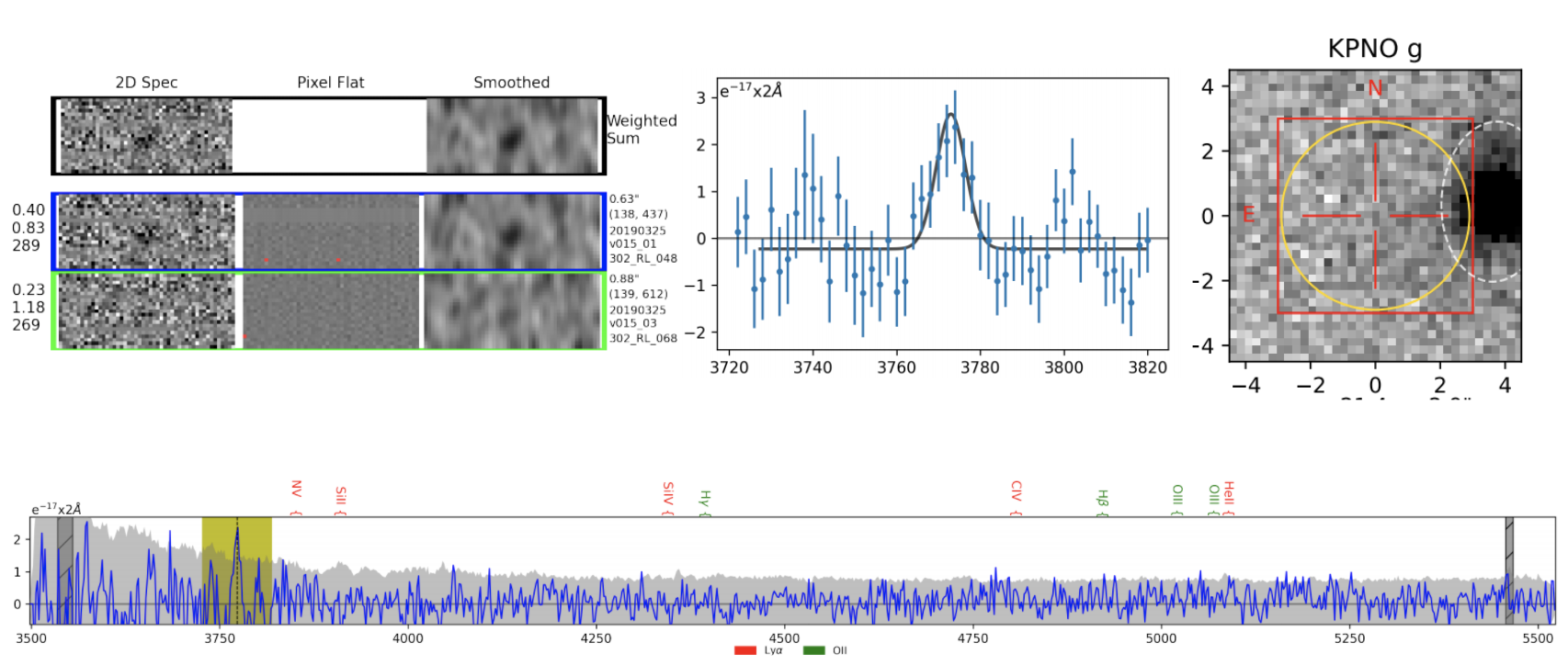}
    \caption{\textbf{20190325v015-9900150001.} \emph{Top Left}: Two best VIRUS fibers, pixel flats, smoothed VIRUS fibers, and weighted sum of fibers. \emph{Top Middle}: Gaussian fit to emission line. \emph{Top Right}: Best available imaging on location of lensed LAE candidate. \emph{Bottom}: Full spectrum of lensed LAE candidate. The yellow box with a dashed line denotes location of Ly$\alpha$.}
\end{figure*}

\begin{figure*}
    \centering
\includegraphics[ width=\textwidth]{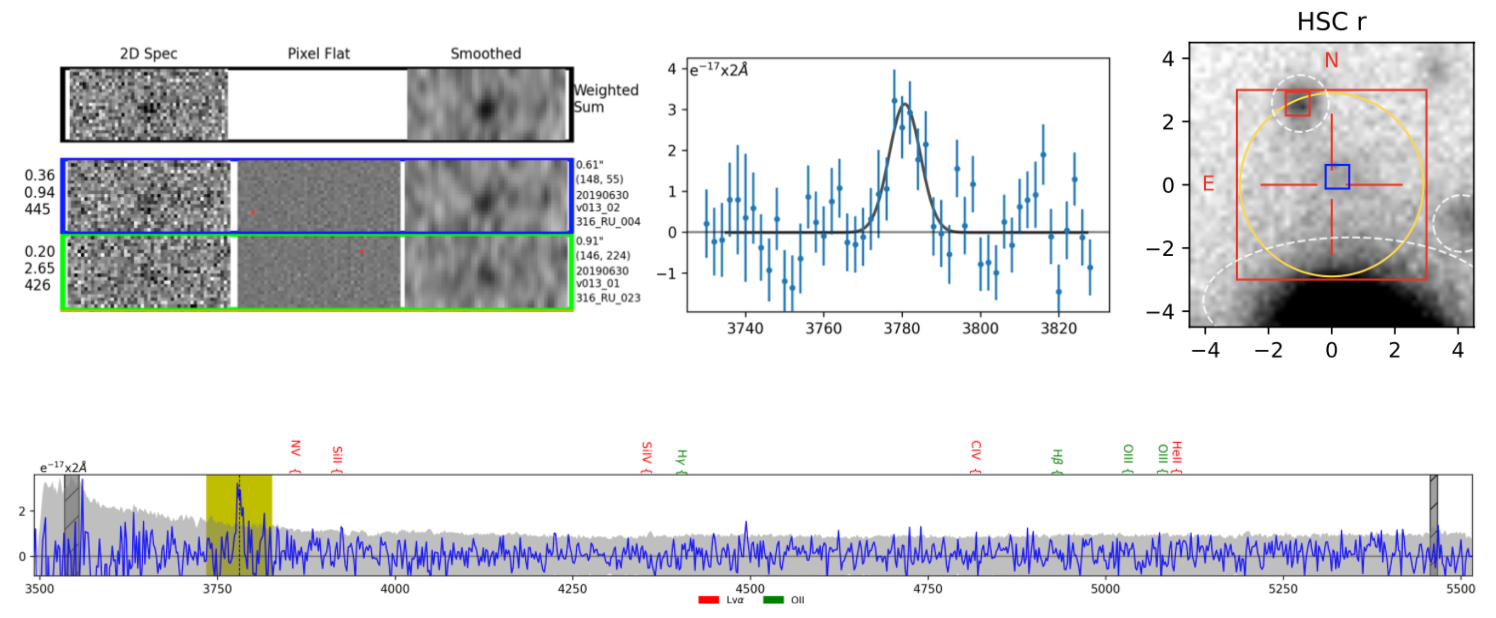}
    \caption{\textbf{20190630v013-9900030001.} \emph{Top Left}: Two best VIRUS fibers, pixel flats, smoothed VIRUS fibers, and weighted sum of fibers. \emph{Top Middle}: Gaussian fit to emission line. \emph{Top Right}: Best available imaging on location of lensed LAE candidate. \emph{Bottom}: Full spectrum of lensed LAE candidate. The yellow box with a dashed line denotes location of Ly$\alpha$.}
\end{figure*}

\begin{figure*}
    \centering
\includegraphics[ width=\textwidth]{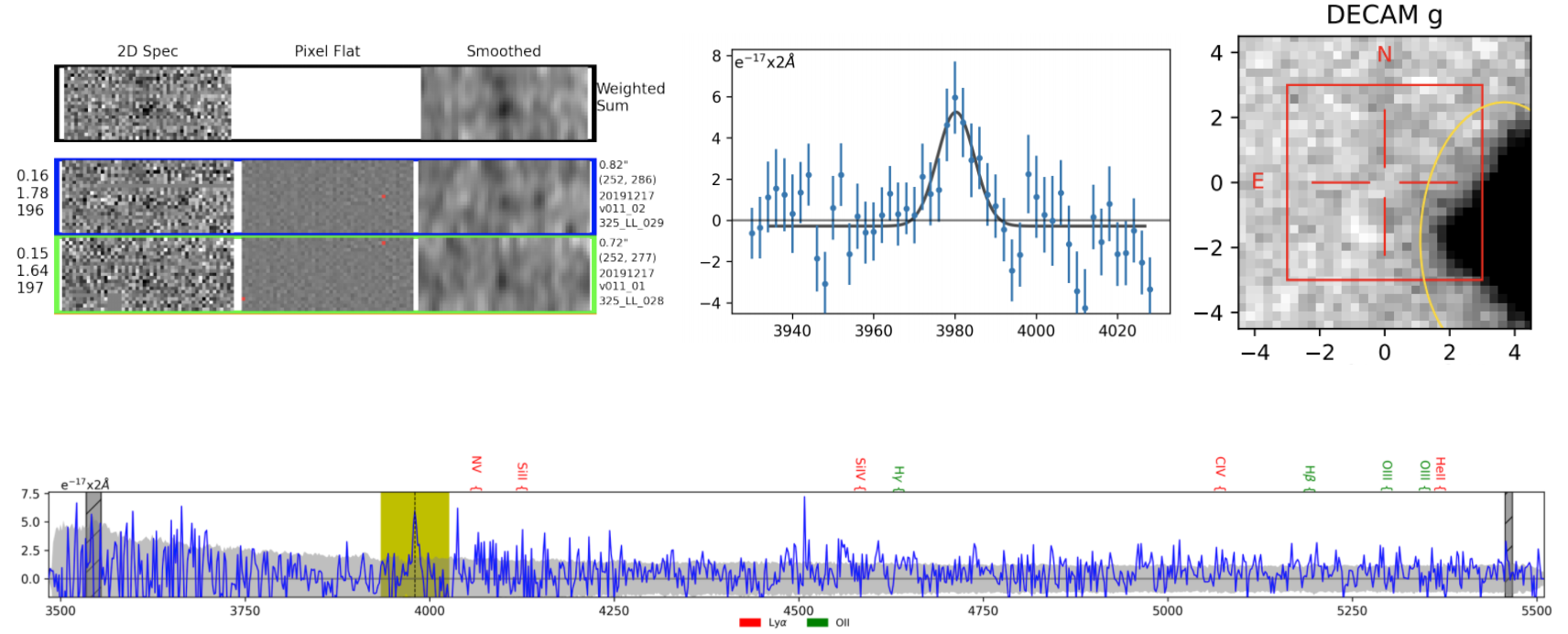}
    \caption{\textbf{20191217v011-9900060004.} \emph{Top Left}: Two best VIRUS fibers, pixel flats, smoothed VIRUS fibers, and weighted sum of fibers. \emph{Top Middle}: Gaussian fit to emission line. \emph{Top Right}: Best available imaging on location of lensed LAE candidate. \emph{Bottom}: Full spectrum of lensed LAE candidate. The yellow box with a dashed line denotes location of Ly$\alpha$.}
\end{figure*}

\begin{figure*}
    \centering
\includegraphics[ width=\textwidth]{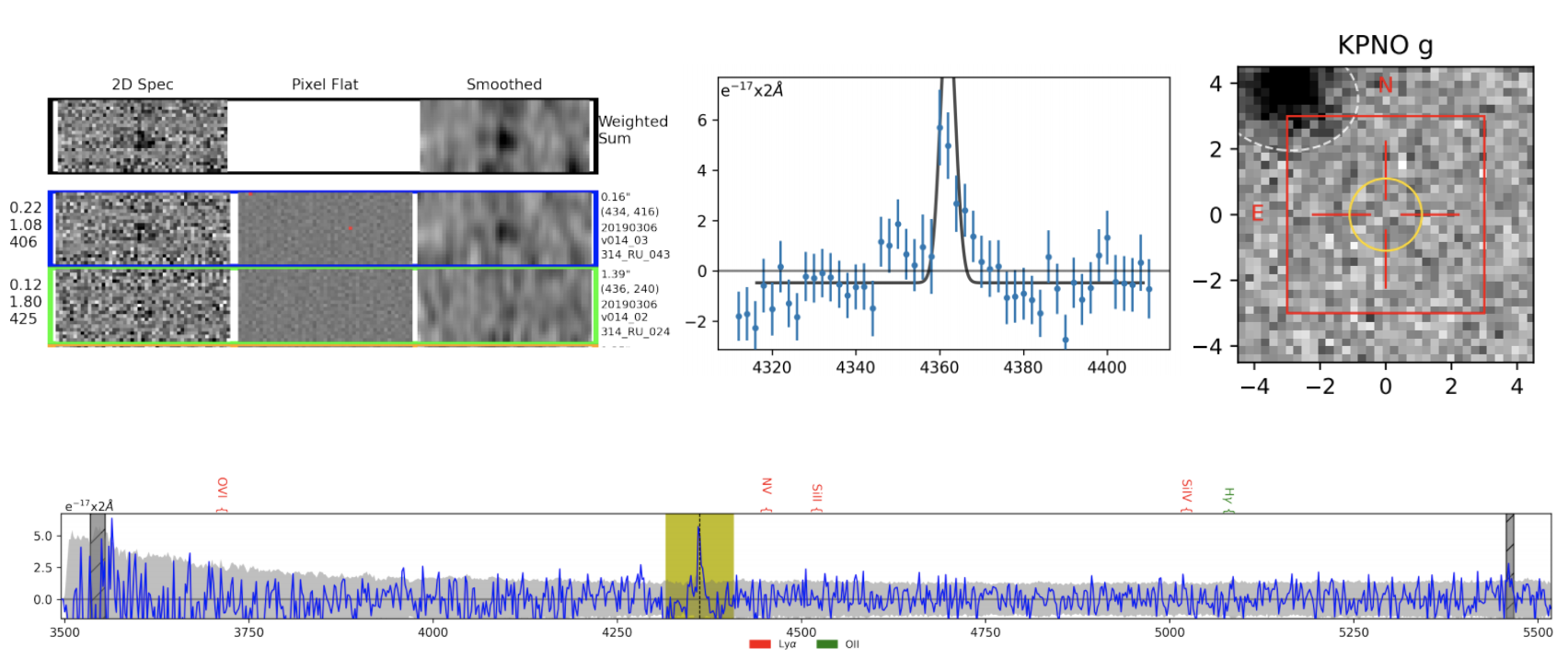}
    \caption{\textbf{20190306v014-9900010001.} \emph{Top Left}: Two best VIRUS fibers, pixel flats, smoothed VIRUS fibers, and weighted sum of fibers. \emph{Top Middle}: Gaussian fit to emission line. \emph{Top Right}: Best available imaging on location of lensed LAE candidate. \emph{Bottom}: Full spectrum of lensed LAE candidate. The yellow box with a dashed line denotes location of Ly$\alpha$.}
\end{figure*}

\begin{figure*}
    \centering
\includegraphics[ width=\textwidth]{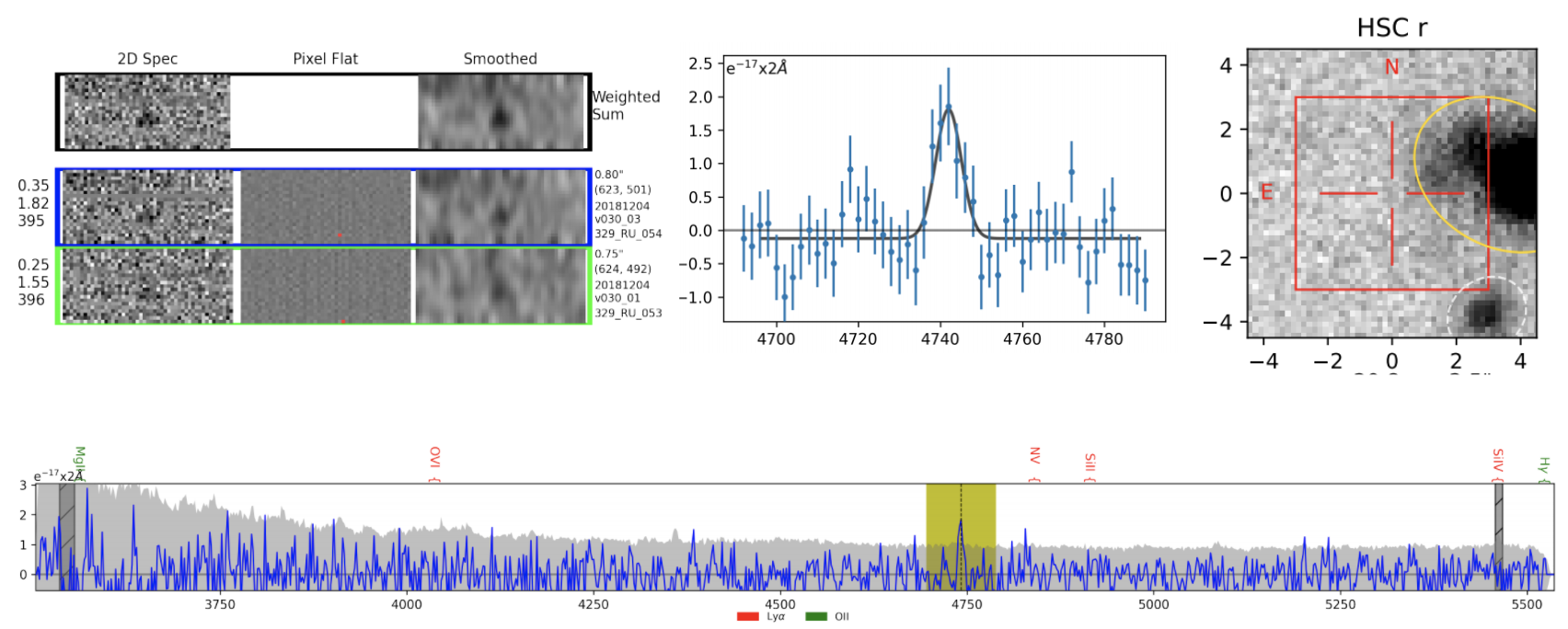}
    \caption{\textbf{20181204v030-9900010004.} \emph{Top Left}: Two best VIRUS fibers, pixel flats, smoothed VIRUS fibers, and weighted sum of fibers. \emph{Top Middle}: Gaussian fit to emission line. \emph{Top Right}: Best available imaging on location of lensed LAE candidate. \emph{Bottom}: Full spectrum of lensed LAE candidate. The yellow box with a dashed line denotes location of Ly$\alpha$.}
\end{figure*}

\begin{figure*}
    \centering
\includegraphics[ width=\textwidth]{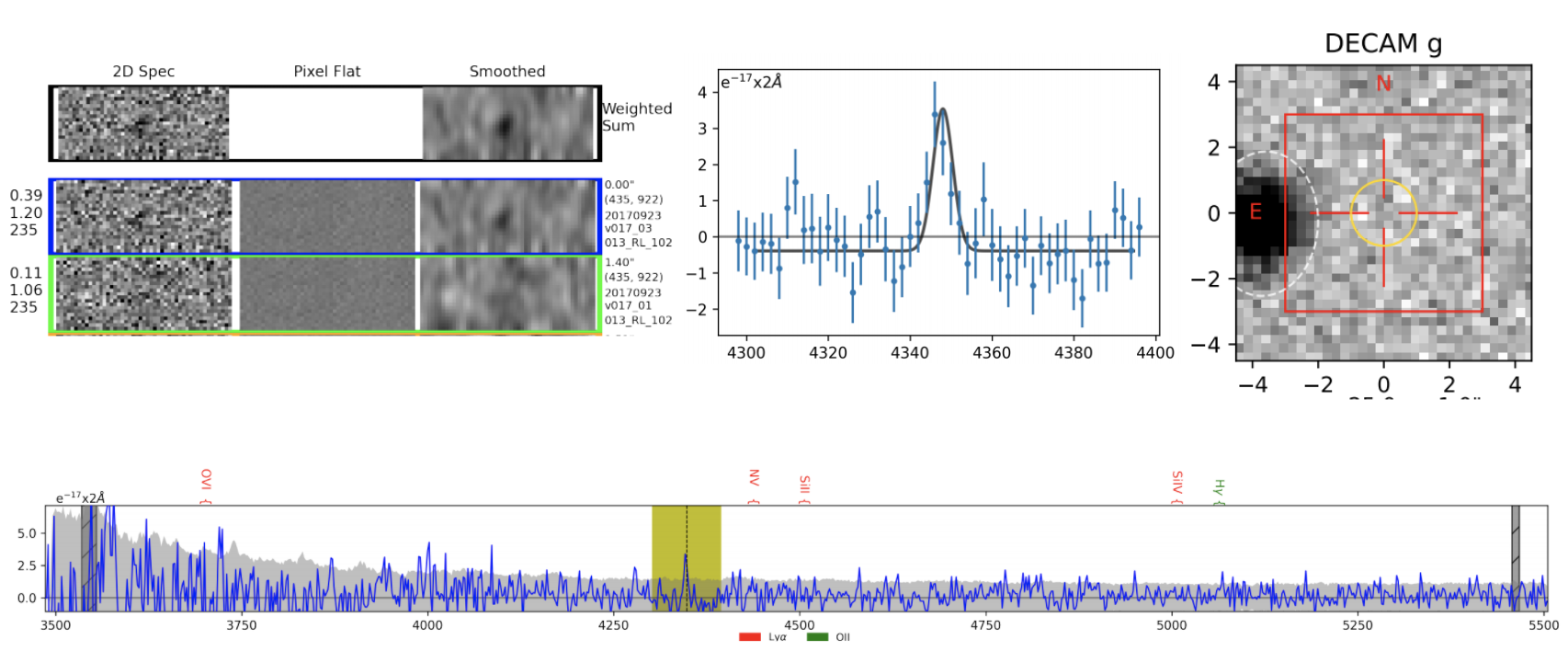}
    \caption{\textbf{20170923v017-9900050003.} \emph{Top Left}: Two best VIRUS fibers, pixel flats, smoothed VIRUS fibers, and weighted sum of fibers. \emph{Top Middle}: Gaussian fit to emission line. \emph{Top Right}: Best available imaging on location of lensed LAE candidate. \emph{Bottom}: Full spectrum of lensed LAE candidate. The yellow box with a dashed line denotes location of Ly$\alpha$.}
\end{figure*}

\begin{figure*}
    \centering
\includegraphics[ width=\textwidth]{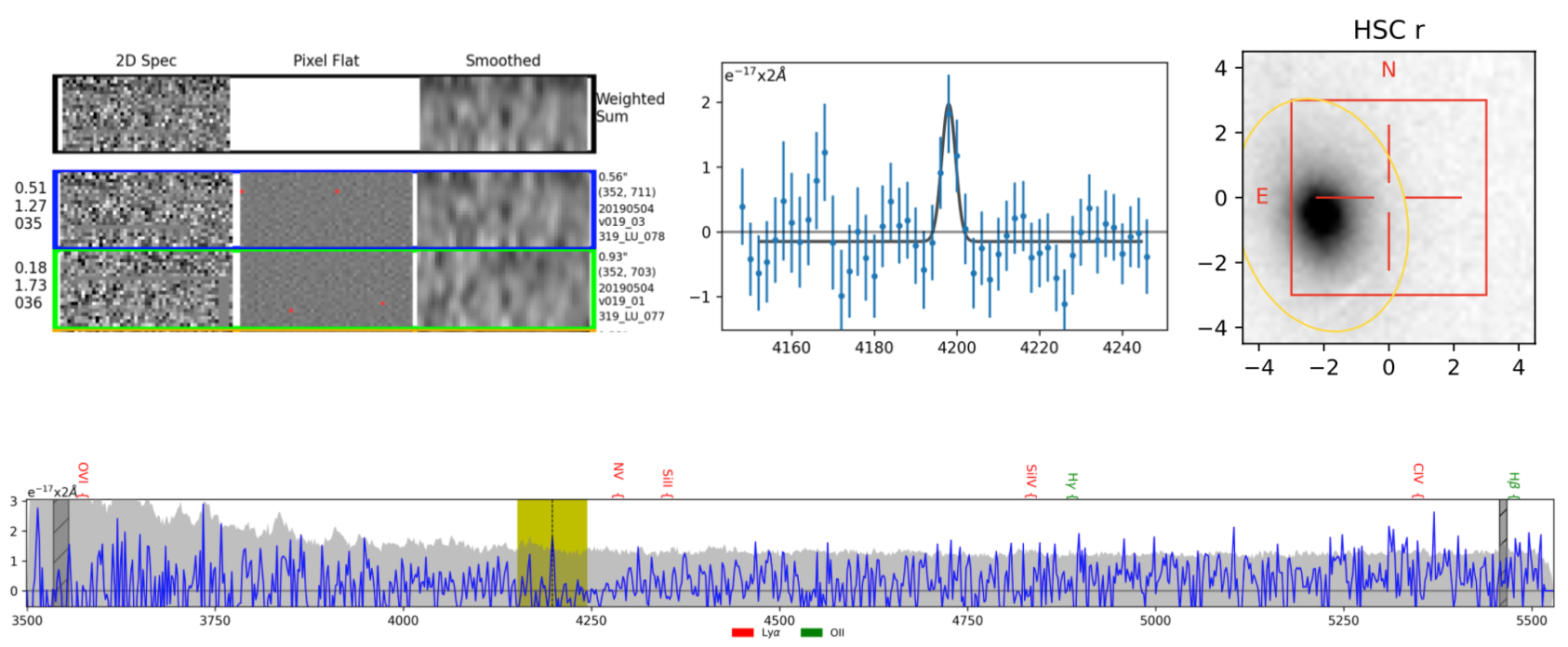}
    \caption{\textbf{20190504v019-9900150006.} \emph{Top Left}: Two best VIRUS fibers, pixel flats, smoothed VIRUS fibers, and weighted sum of fibers. \emph{Top Middle}: Gaussian fit to emission line. \emph{Top Right}: Best available imaging on location of lensed LAE candidate. \emph{Bottom}: Full spectrum of lensed LAE candidate. The yellow box with a dashed line denotes location of Ly$\alpha$.}
\end{figure*}

\begin{figure*}
    \centering
\includegraphics[ width=\textwidth]{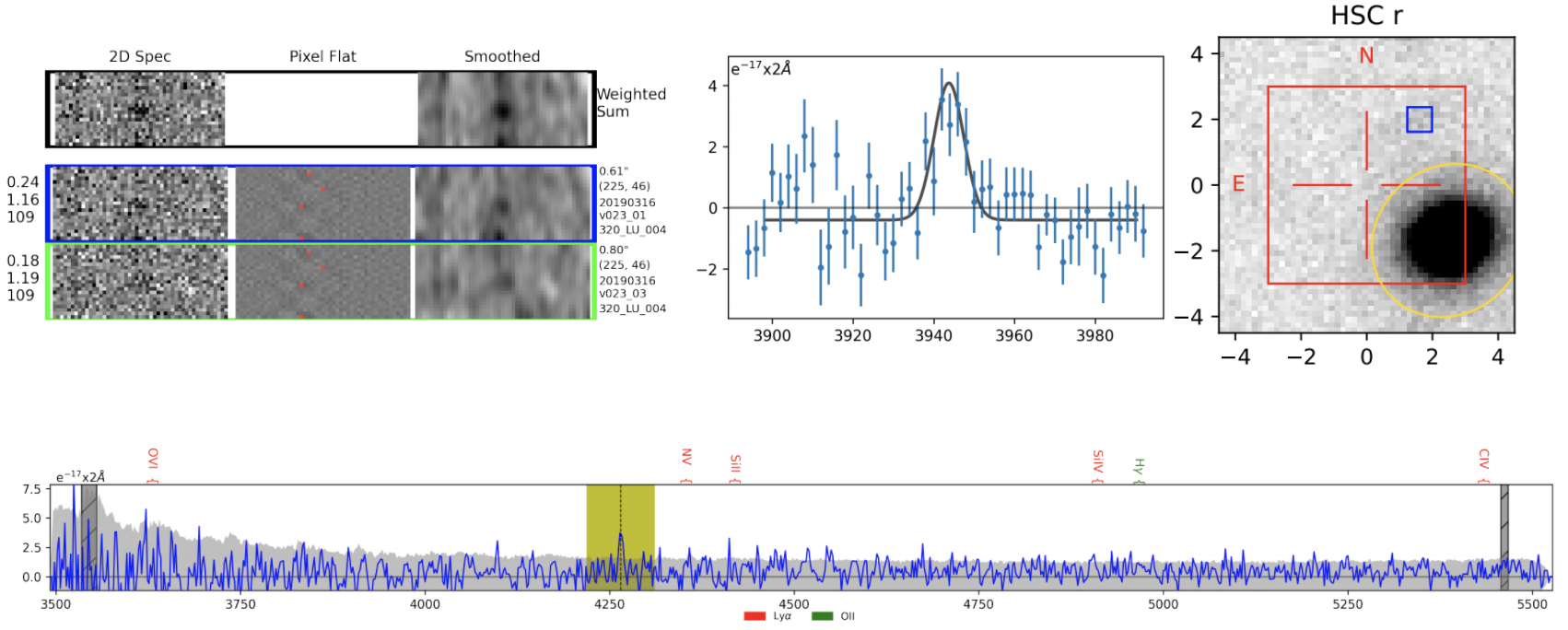}
    \caption{\textbf{20190316v023-9900090004.} \emph{Top Left}: Two best VIRUS fibers, pixel flats, smoothed VIRUS fibers, and weighted sum of fibers. \emph{Top Middle}: Gaussian fit to emission line. \emph{Top Right}: Best available imaging on location of lensed LAE candidate. \emph{Bottom}: Full spectrum of lensed LAE candidate. The yellow box with a dashed line denotes location of Ly$\alpha$.}
\end{figure*}

\begin{figure*}
    \centering
\includegraphics[ width=\textwidth]{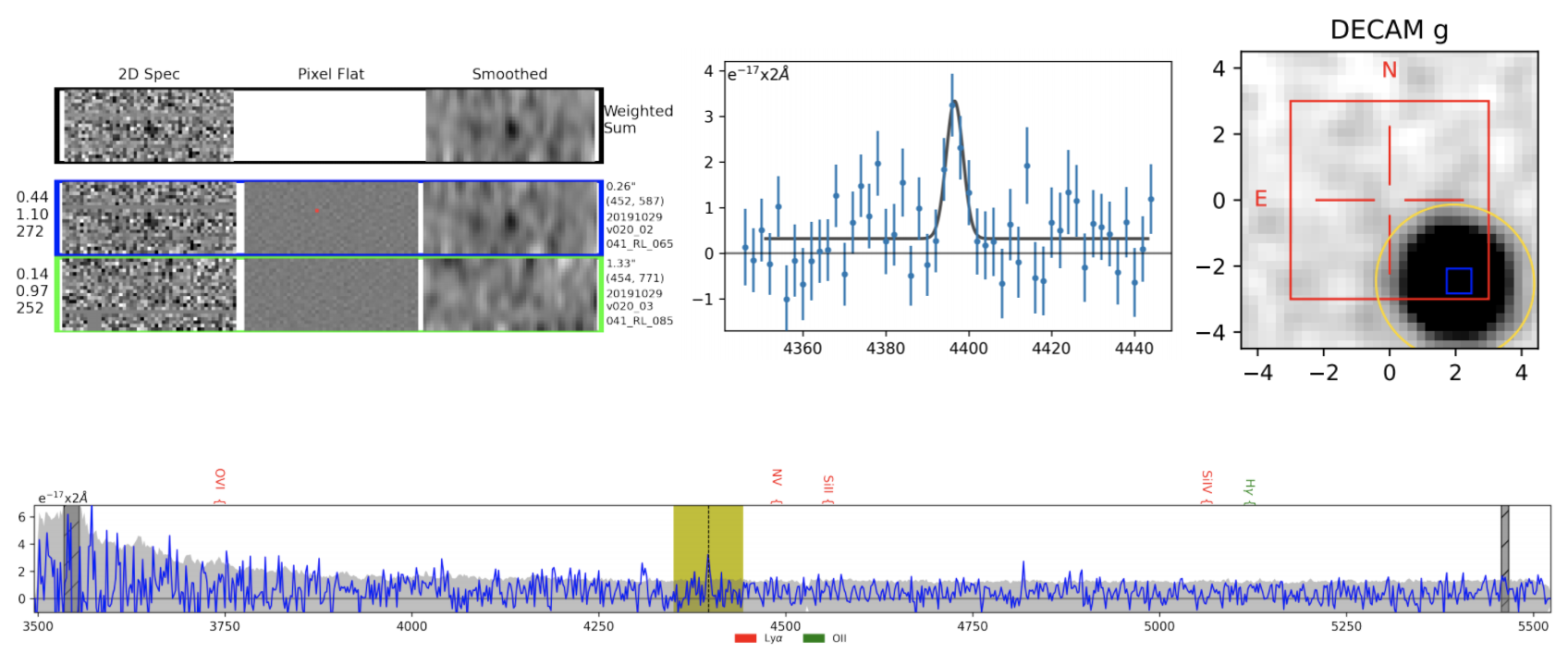}
    \caption{\textbf{20191029v020-9900140002.} \emph{Top Left}: Two best VIRUS fibers, pixel flats, smoothed VIRUS fibers, and weighted sum of fibers. \emph{Top Middle}: Gaussian fit to emission line. \emph{Top Right}: Best available imaging on location of lensed LAE candidate. \emph{Bottom}: Full spectrum of lensed LAE candidate. The yellow box with a dashed line denotes location of Ly$\alpha$.}
\end{figure*}

\begin{figure*}
    \centering
\includegraphics[ width=\textwidth]{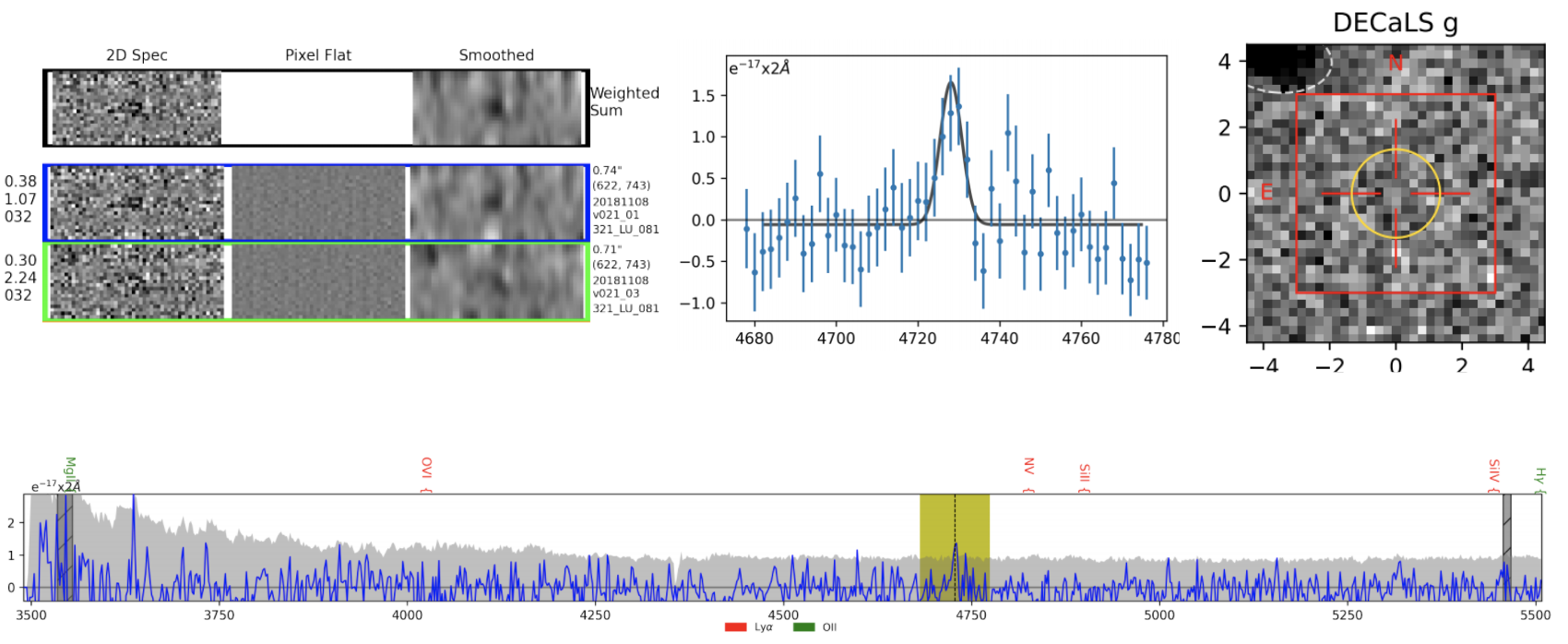}
    \caption{\textbf{20181108v021-9900080002.} \emph{Top Left}: Two best VIRUS fibers, pixel flats, smoothed VIRUS fibers, and weighted sum of fibers. \emph{Top Middle}: Gaussian fit to emission line. \emph{Top Right}: Best available imaging on location of lensed LAE candidate. \emph{Bottom}: Full spectrum of lensed LAE candidate. The yellow box with a dashed line denotes location of Ly$\alpha$.}
\end{figure*}

\begin{figure*}
    \centering
\includegraphics[ width=\textwidth]{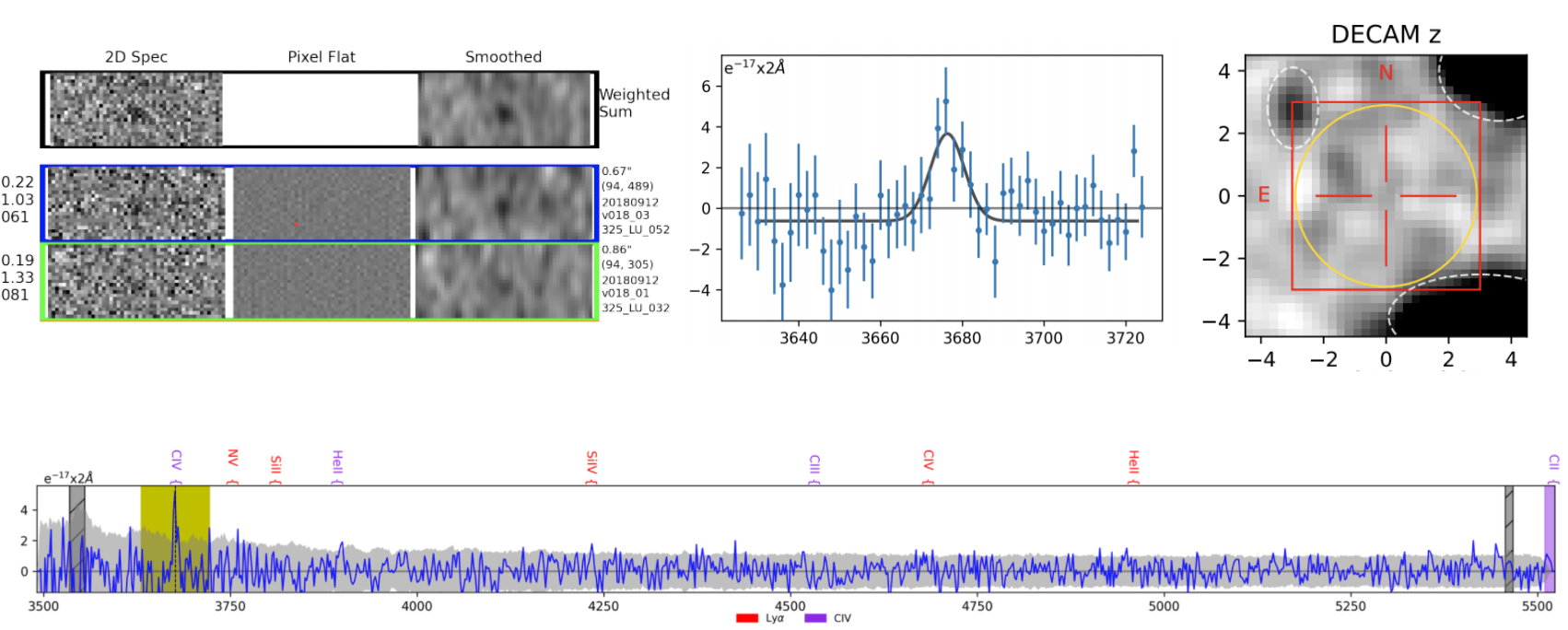}
    \caption{\textbf{20180912v018-9900070002.} \emph{Top Left}: Two best VIRUS fibers, pixel flats, smoothed VIRUS fibers, and weighted sum of fibers. \emph{Top Middle}: Gaussian fit to emission line. \emph{Top Right}: Best available imaging on location of lensed LAE candidate. \emph{Bottom}: Full spectrum of lensed LAE candidate. The yellow box with a dashed line denotes location of Ly$\alpha$.}
\end{figure*}

\begin{figure*}
    \centering
\includegraphics[ width=\textwidth]{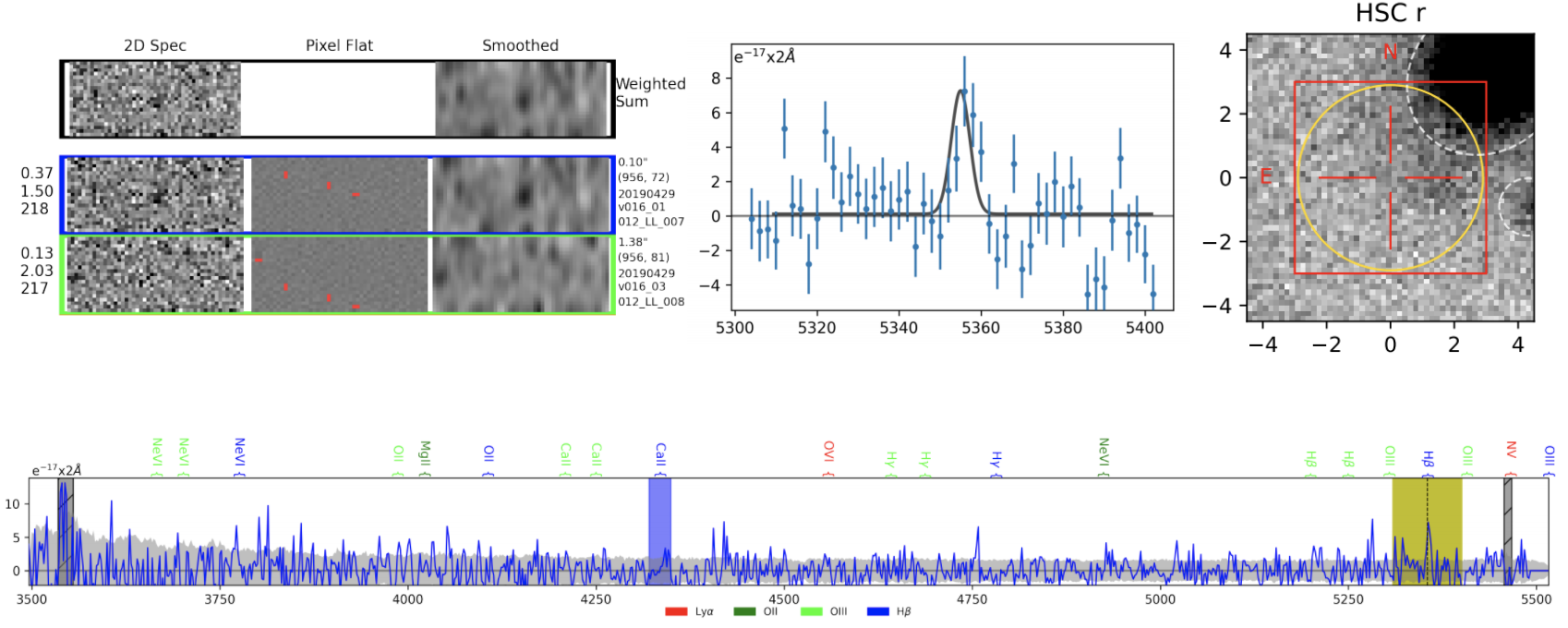}
    \caption{\textbf{20190429v016-9900140004.} \emph{Top Left}: Two best VIRUS fibers, pixel flats, smoothed VIRUS fibers, and weighted sum of fibers. \emph{Top Middle}: Gaussian fit to emission line. \emph{Top Right}: Best available imaging on location of lensed LAE candidate. \emph{Bottom}: Full spectrum of lensed LAE candidate. The yellow box with a dashed line denotes location of Ly$\alpha$.}
\end{figure*}

\begin{figure*}
    \centering
\includegraphics[ width=\textwidth]{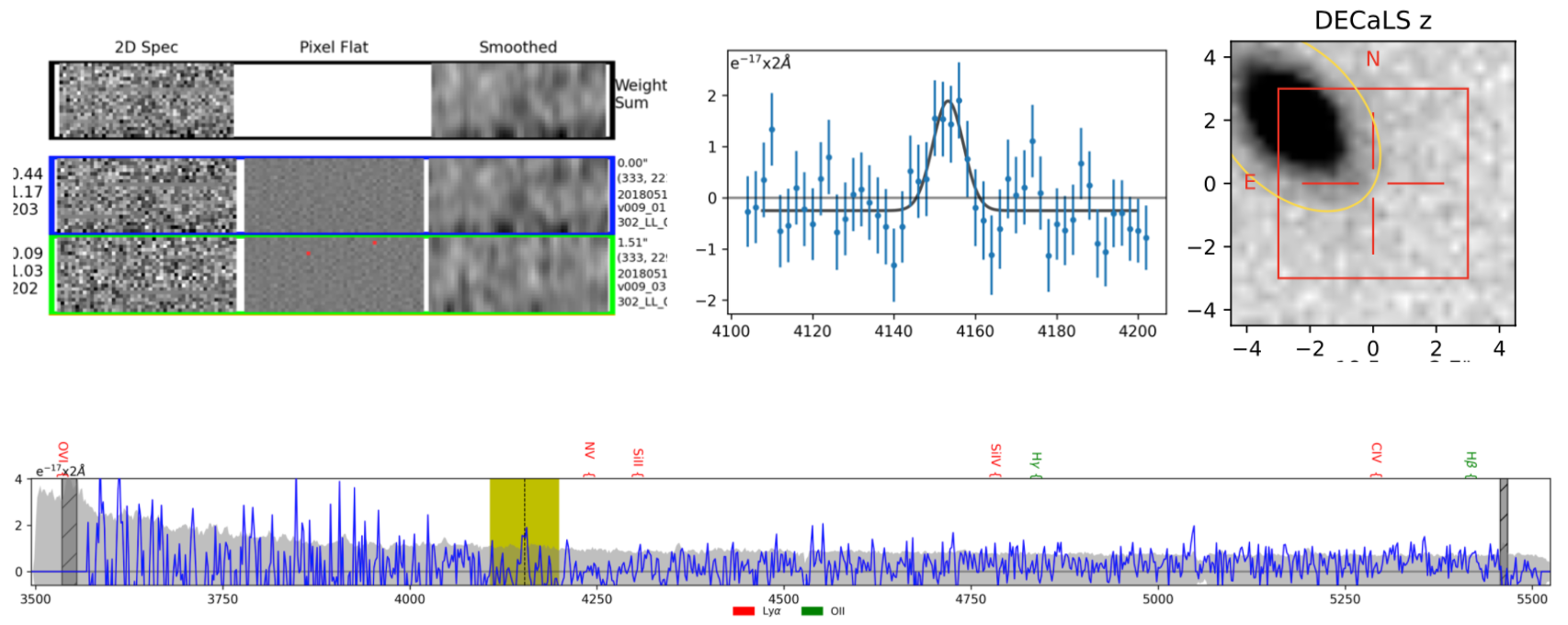}
    \caption{\textbf{20180516v009-9900100002.} \emph{Top Left}: Two best VIRUS fibers, pixel flats, smoothed VIRUS fibers, and weighted sum of fibers. \emph{Top Middle}: Gaussian fit to emission line. \emph{Top Right}: Best available imaging on location of lensed LAE candidate. \emph{Bottom}: Full spectrum of lensed LAE candidate. The yellow box with a dashed line denotes location of Ly$\alpha$.}
\end{figure*}

\begin{figure*}
    \centering
\includegraphics[ width=\textwidth]{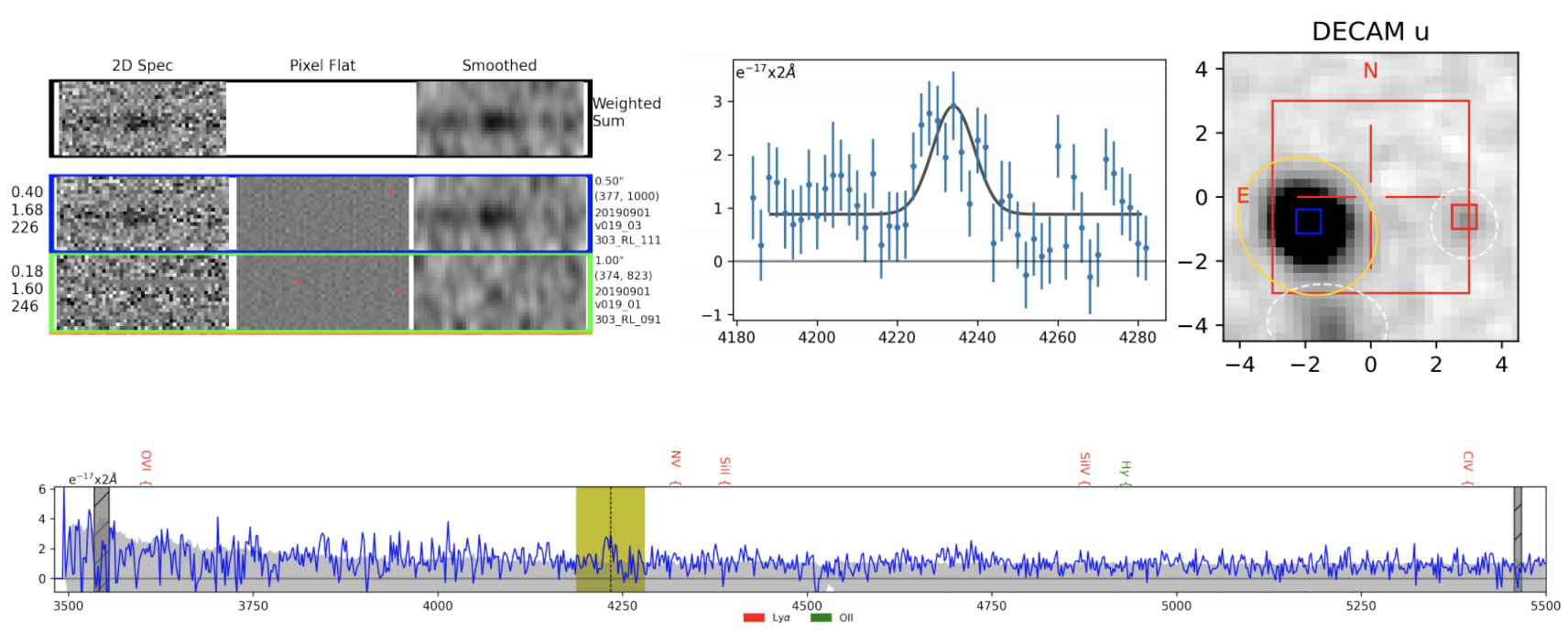}
    \caption{\textbf{20190901v019-9900190001.} \emph{Top Left}: Two best VIRUS fibers, pixel flats, smoothed VIRUS fibers, and weighted sum of fibers. \emph{Top Middle}: Gaussian fit to emission line. \emph{Top Right}: Best available imaging on location of lensed LAE candidate. \emph{Bottom}: Full spectrum of lensed LAE candidate. The yellow box with a dashed line denotes location of Ly$\alpha$.}
\end{figure*}

\begin{figure*}
    \centering
\includegraphics[ width=\textwidth]{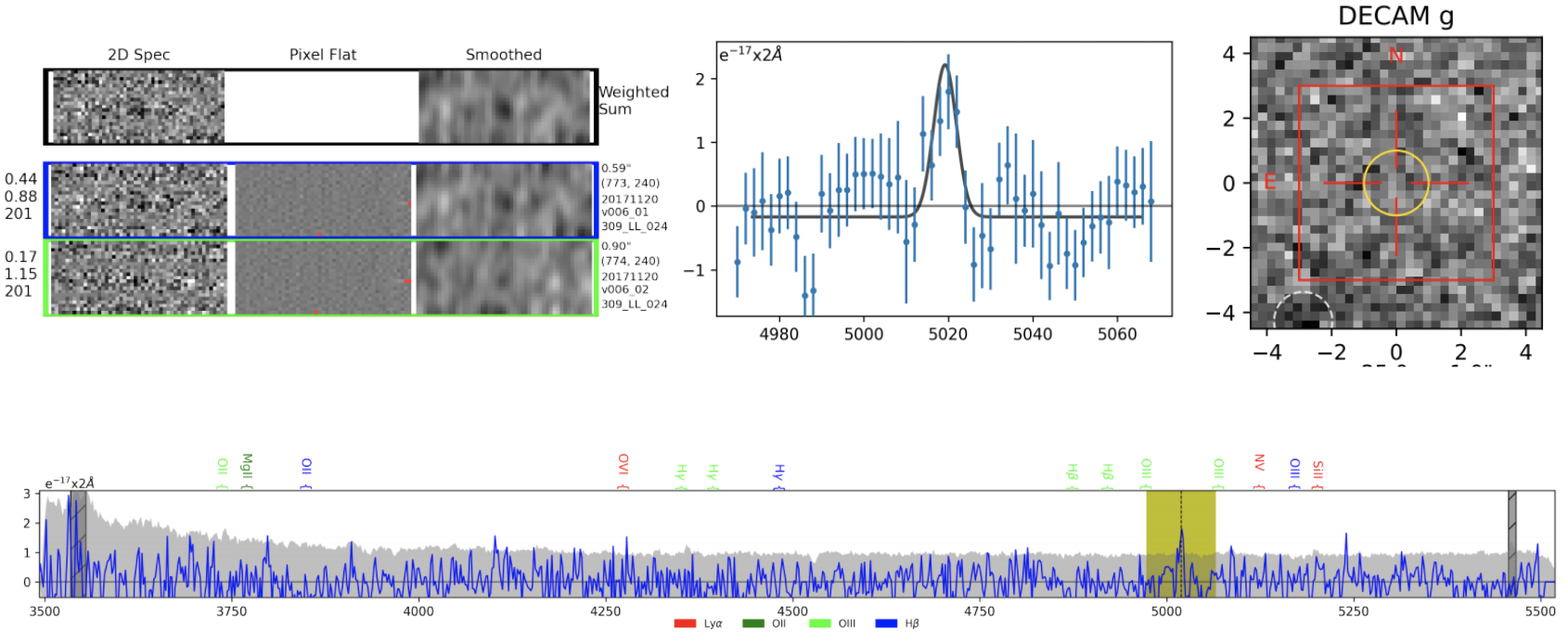}
    \caption{\textbf{20171120v006-9900050002.} \emph{Top Left}: Two best VIRUS fibers, pixel flats, smoothed VIRUS fibers, and weighted sum of fibers. \emph{Top Middle}: Gaussian fit to emission line. \emph{Top Right}: Best available imaging on location of lensed LAE candidate. \emph{Bottom}: Full spectrum of lensed LAE candidate. The yellow box with a dashed line denotes location of Ly$\alpha$.}
\end{figure*}

\begin{figure*}
    \centering
\includegraphics[ width=\textwidth]{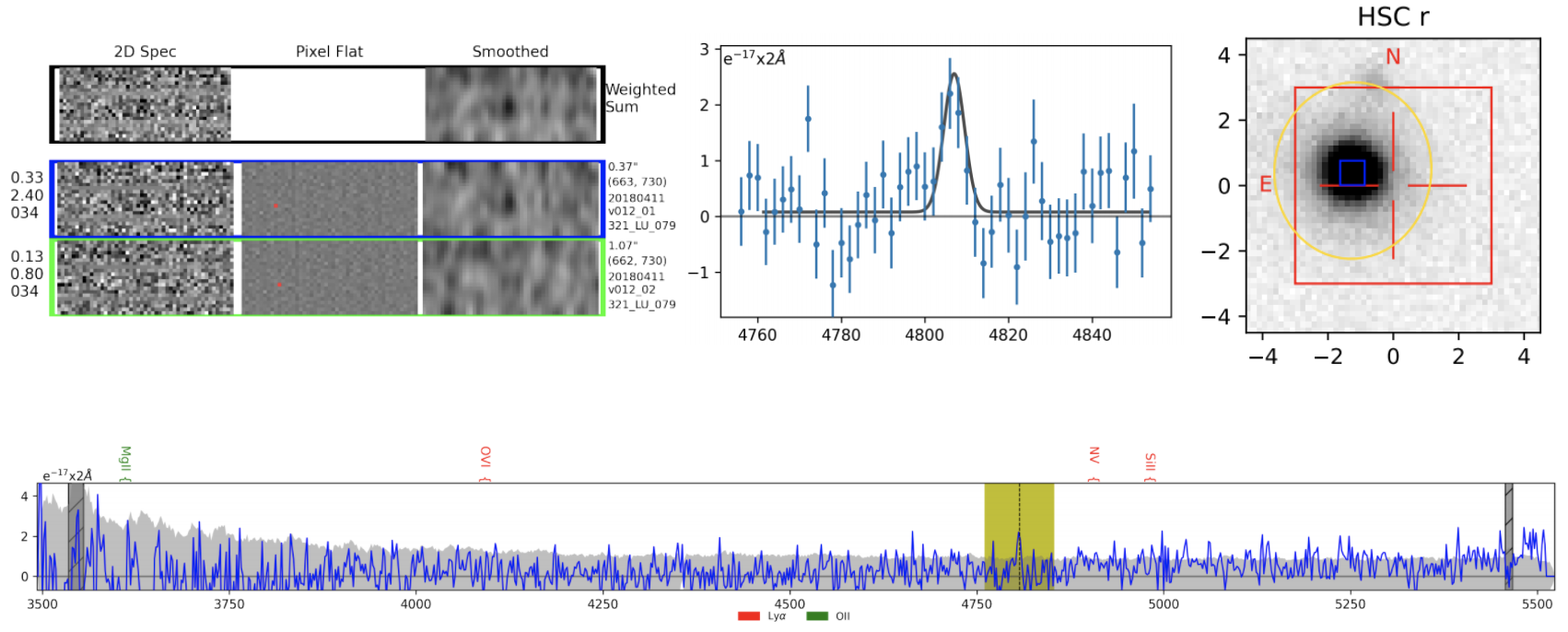}
    \caption{\textbf{20180411v012-9900160001.} \emph{Top Left}: Two best VIRUS fibers, pixel flats, smoothed VIRUS fibers, and weighted sum of fibers. \emph{Top Middle}: Gaussian fit to emission line. \emph{Top Right}: Best available imaging on location of lensed LAE candidate. \emph{Bottom}: Full spectrum of lensed LAE candidate. The yellow box with a dashed line denotes location of Ly$\alpha$.}
    \label{appendix fig 2}
\end{figure*}

\begin{figure*}
    \centering
\includegraphics[ width=\textwidth]{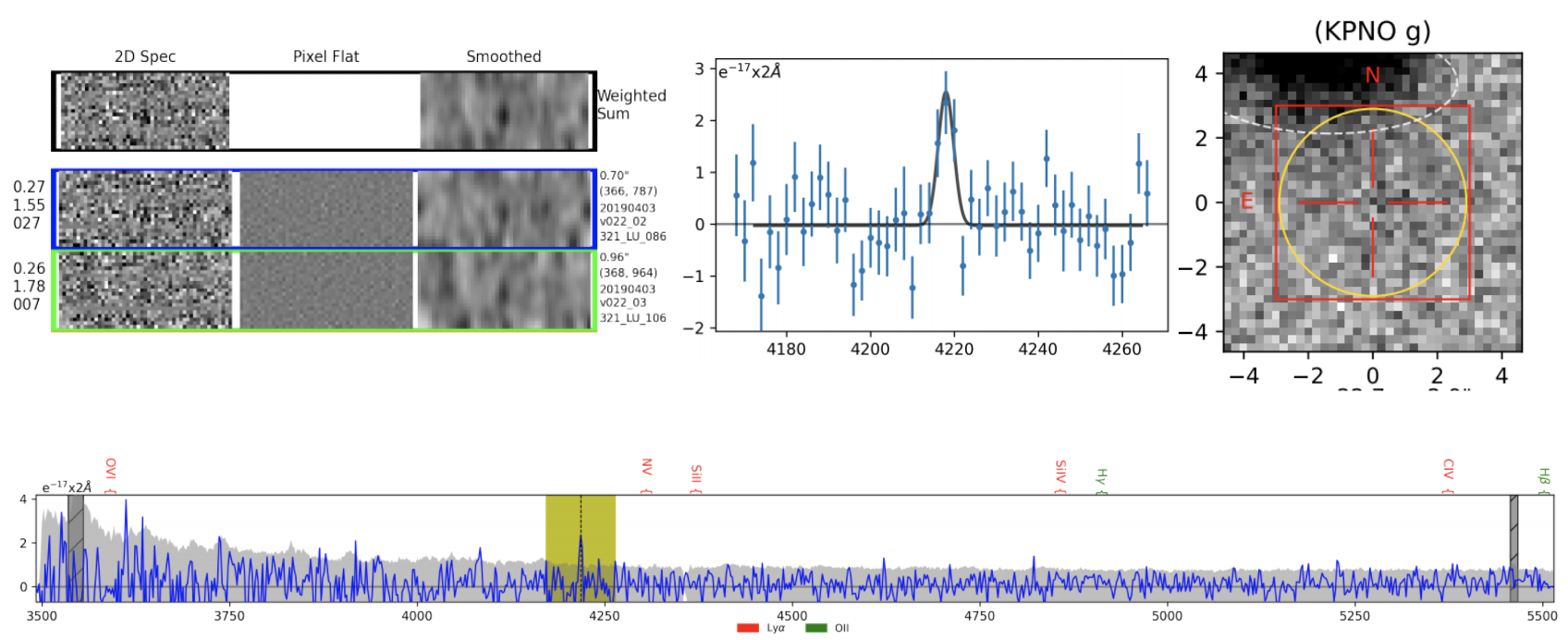}
    \caption{\textbf{20190403v022-9900000002.} \emph{Top Left}: Two best VIRUS fibers, pixel flats, smoothed VIRUS fibers, and weighted sum of fibers. \emph{Top Middle}: Gaussian fit to emission line. \emph{Top Right}: Best available imaging on location of lensed LAE candidate. \emph{Bottom}: Full spectrum of lensed LAE candidate. The yellow box with a dashed line denotes location of Ly$\alpha$.}
\end{figure*}

\begin{figure*}
    \centering
\includegraphics[ width=\textwidth]{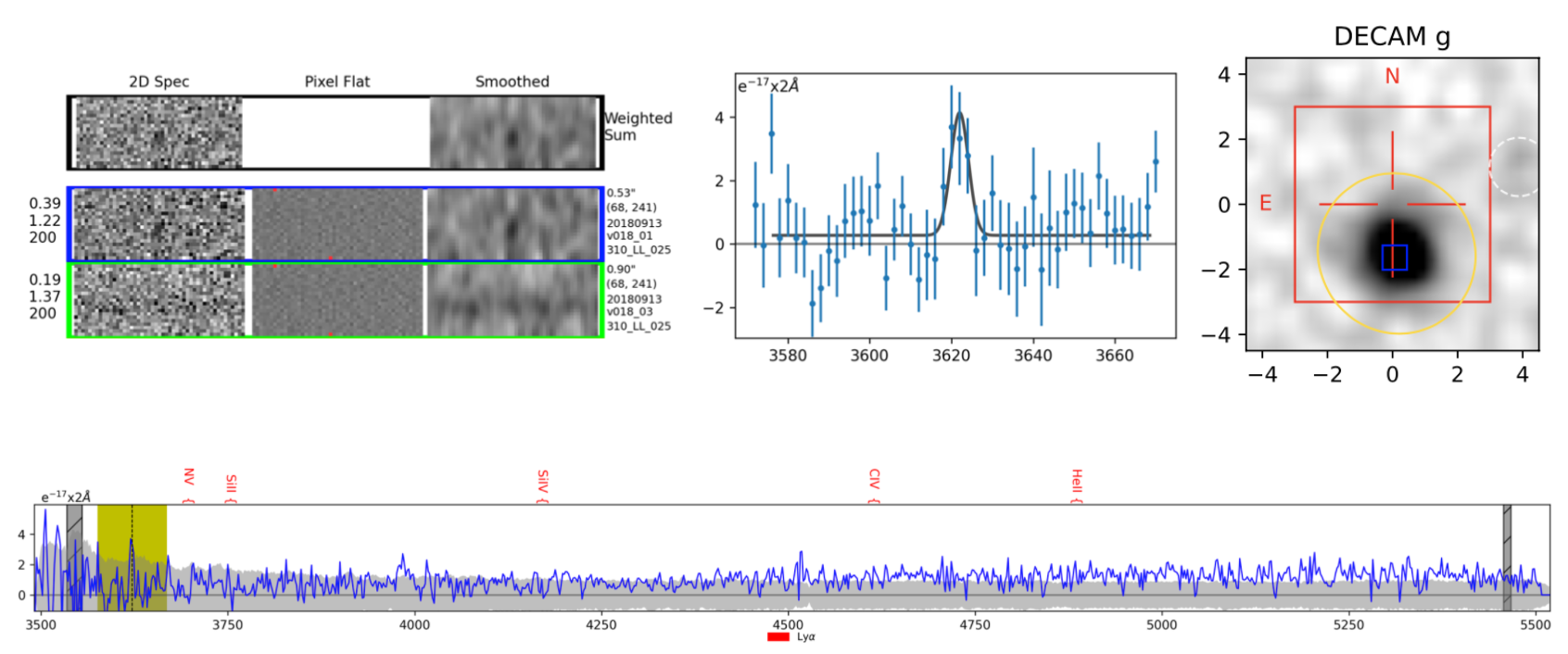}
    \caption{\textbf{20180913v018-9900100003.} \emph{Top Left}: Two best VIRUS fibers, pixel flats, smoothed VIRUS fibers, and weighted sum of fibers. \emph{Top Middle}: Gaussian fit to emission line. \emph{Top Right}: Best available imaging on location of lensed LAE candidate. \emph{Bottom}: Full spectrum of lensed LAE candidate. The yellow box with a dashed line denotes location of Ly$\alpha$.}
\end{figure*}

\begin{figure*}
    \centering
\includegraphics[ width=\textwidth]{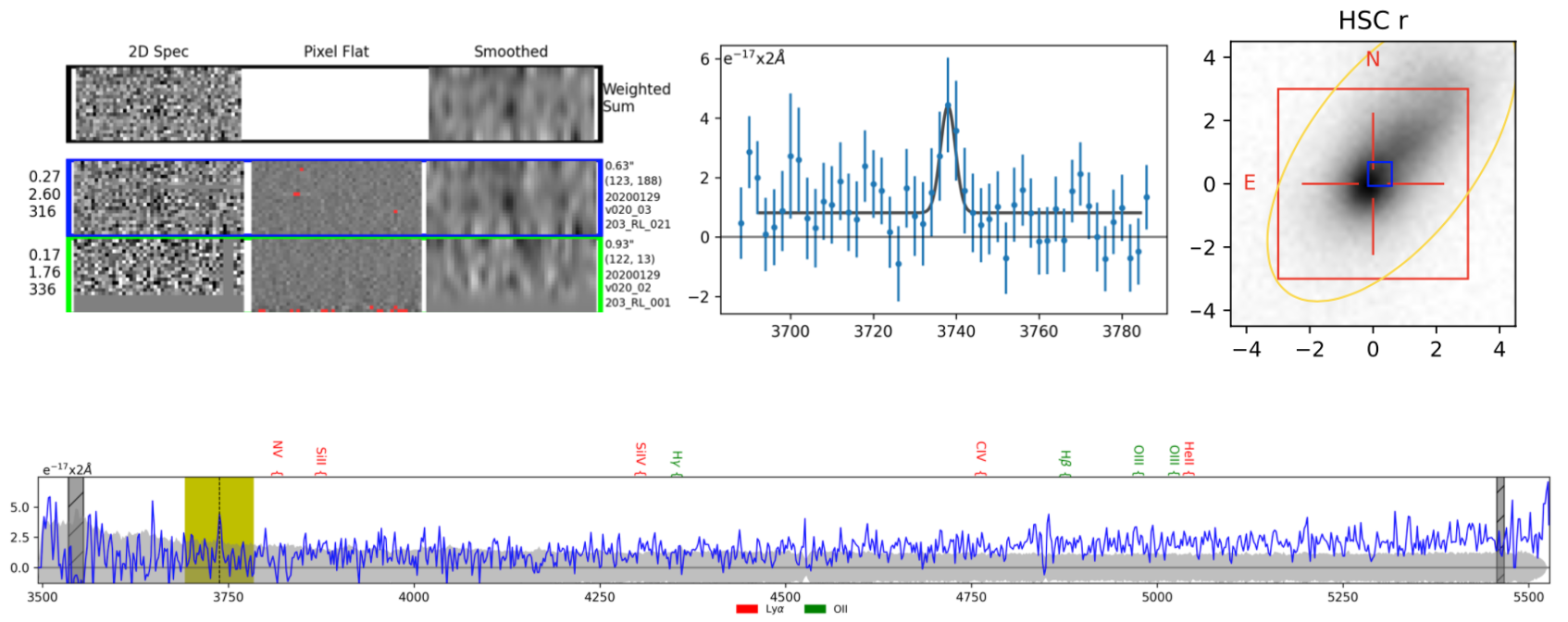}
    \caption{\textbf{20200129v020-9900060001.} \emph{Top Left}: Two best VIRUS fibers, pixel flats, smoothed VIRUS fibers, and weighted sum of fibers. \emph{Top Middle}: Gaussian fit to emission line. \emph{Top Right}: Best available imaging on location of lensed LAE candidate. \emph{Bottom}: Full spectrum of lensed LAE candidate. The yellow box with a dashed line denotes location of Ly$\alpha$.}
\end{figure*}

\begin{figure*}
    \centering
\includegraphics[ width=\textwidth]{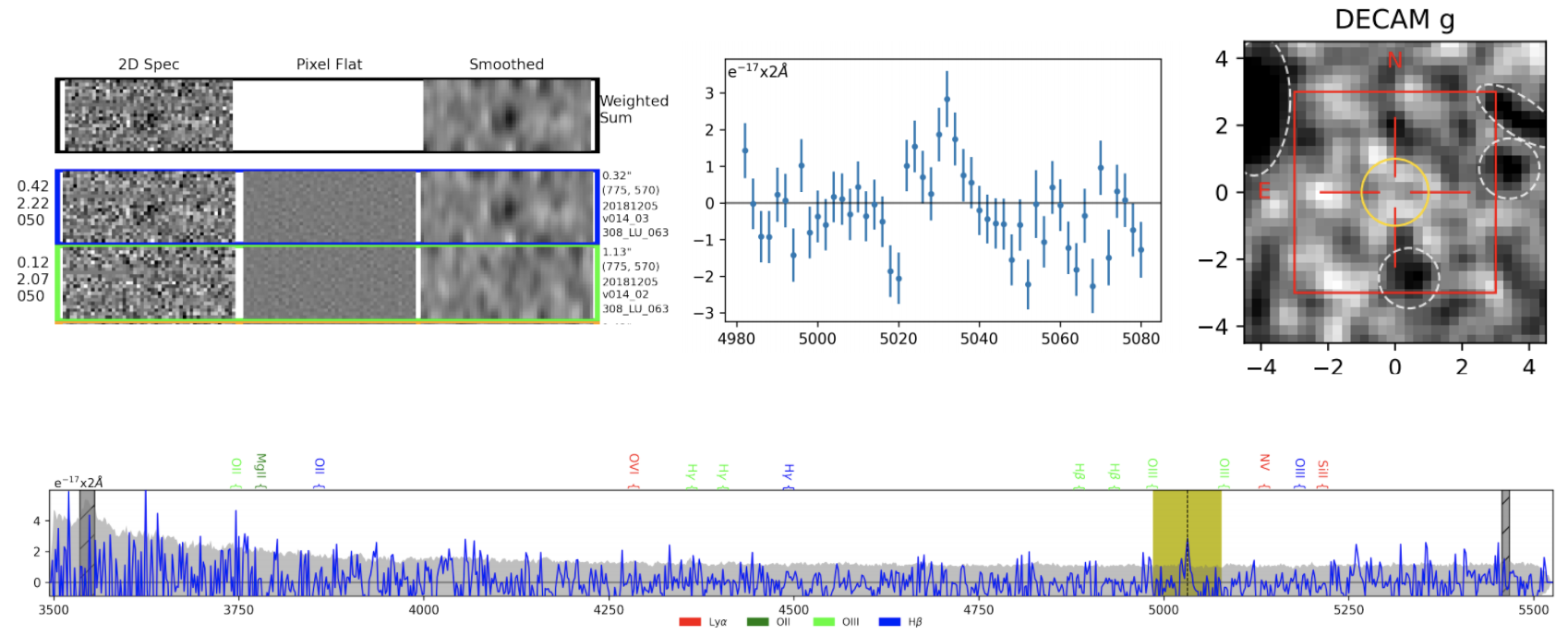}
    \caption{\textbf{20181205v014-9900080003.} \emph{Top Left}: Two best VIRUS fibers, pixel flats, smoothed VIRUS fibers, and weighted sum of fibers. \emph{Top Middle}: Gaussian fit to emission line. \emph{Top Right}: Best available imaging on location of lensed LAE candidate. \emph{Bottom}: Full spectrum of lensed LAE candidate. The yellow box with a dashed line denotes location of Ly$\alpha$.}
\end{figure*}

\begin{figure*}
    \centering
\includegraphics[ width=\textwidth]{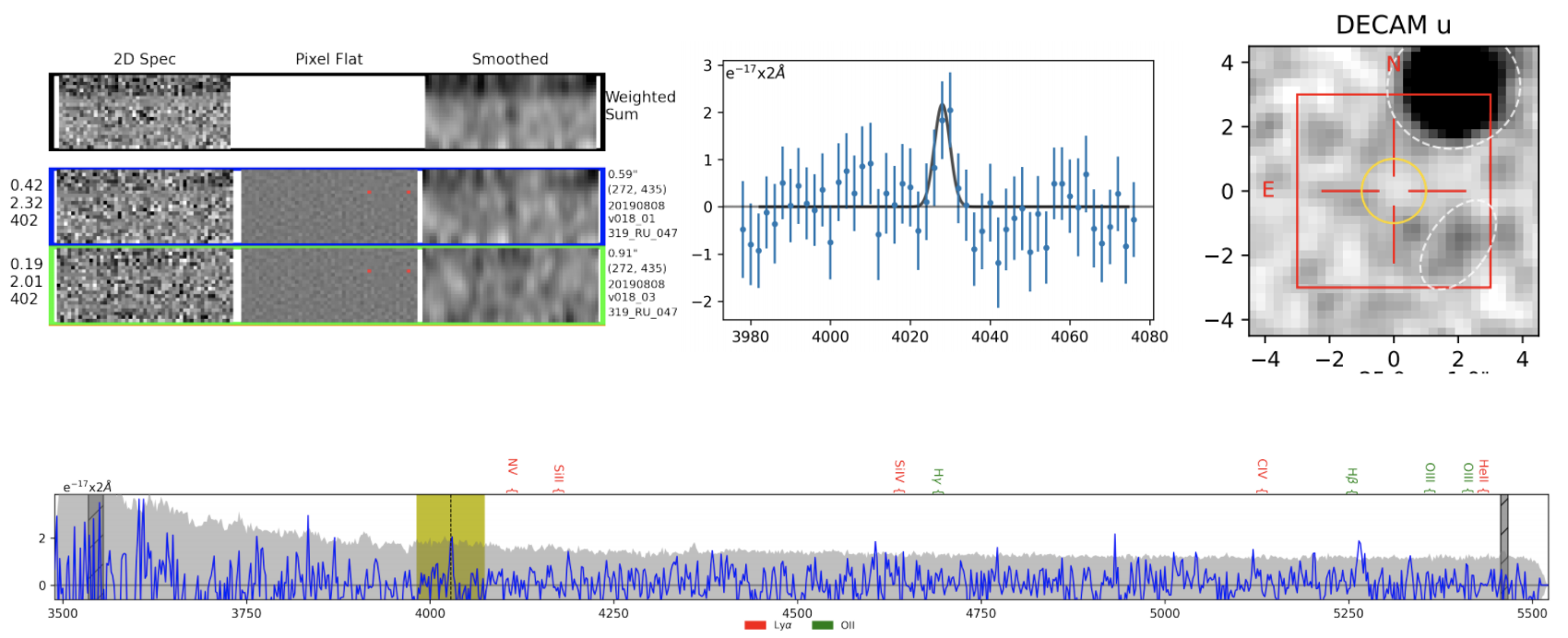}
    \caption{\textbf{20190808v018-9900040003.} \emph{Top Left}: Two best VIRUS fibers, pixel flats, smoothed VIRUS fibers, and weighted sum of fibers. \emph{Top Middle}: Gaussian fit to emission line. \emph{Top Right}: Best available imaging on location of lensed LAE candidate. \emph{Bottom}: Full spectrum of lensed LAE candidate. The yellow box with a dashed line denotes location of Ly$\alpha$.}
\end{figure*}

\begin{figure*}
    \centering
\includegraphics[ width=\textwidth]{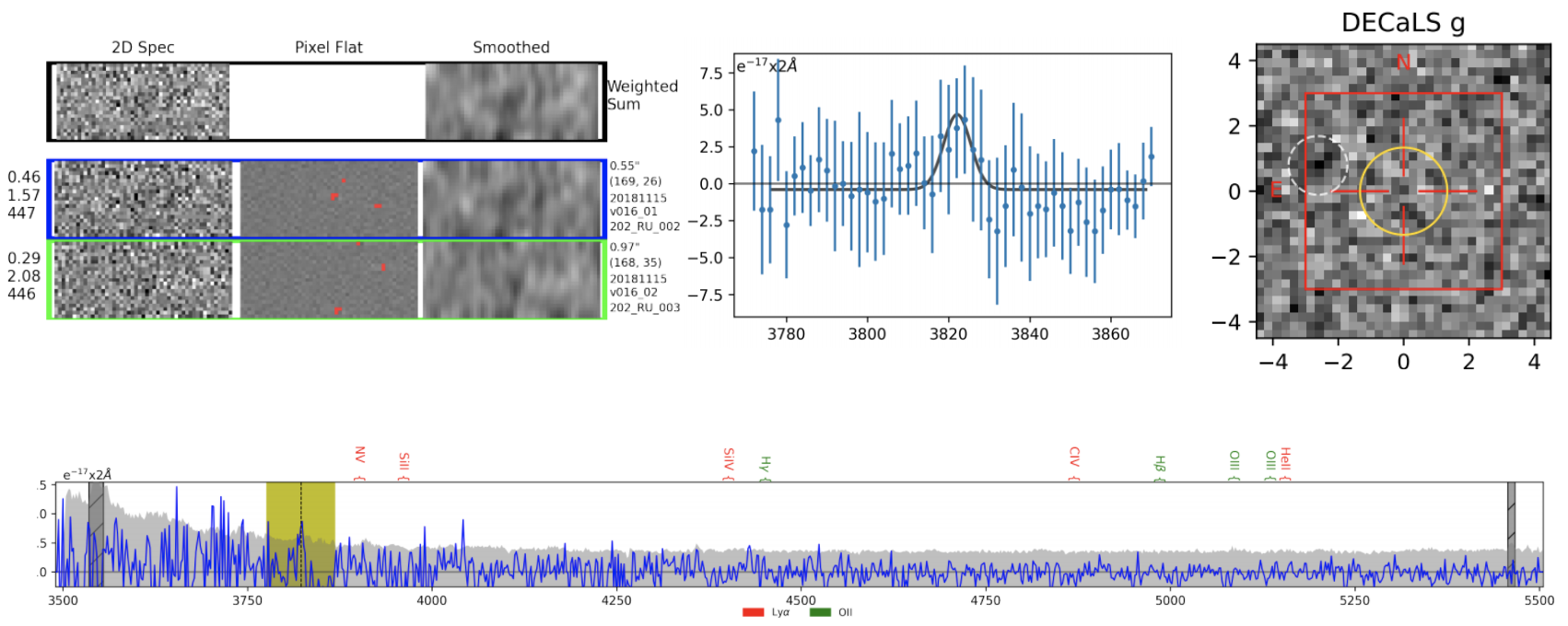}
    \caption{\textbf{20181115v016-9900070005.} \emph{Top Left}: Two best VIRUS fibers, pixel flats, smoothed VIRUS fibers, and weighted sum of fibers. \emph{Top Middle}: Gaussian fit to emission line. \emph{Top Right}: Best available imaging on location of lensed LAE candidate. \emph{Bottom}: Full spectrum of lensed LAE candidate. The yellow box with a dashed line denotes location of Ly$\alpha$.}
    \label{appendix fig 3}
\end{figure*}

\end{document}